\documentclass[runningheads]{llncs}
\usepackage[T1]{fontenc}
\usepackage{subfiles}
\usepackage{graphicx}

\usepackage{xcolor}
\usepackage{amssymb}
\usepackage{amsmath}
\usepackage{algorithm}
\usepackage{algpseudocode}
\usepackage{algorithmicx}%
\usepackage{listings}%
\usepackage[noend]{algcompatible}
\usepackage{hyperref}
\usepackage{multirow}
\usepackage{longtable}

\usepackage{caption}
\usepackage{subcaption}
\definecolor{delim}{RGB}{20,105,176}
\definecolor{numb}{RGB}{106, 109, 32}
\definecolor{string}{rgb}{0.64,0.08,0.08}

\begin{document}
%
\title{A Novel Approach to Process Discovery with Enhanced Loop Handling\\(Extended version)}

\author{Ali {Nour Eldin}\inst{1,2} \and
Benjamin Dalmas\inst{2} \and
Walid Gaaloul\inst{1}}

\authorrunning{A. Nour Eldin et al.}
\institute{Telecom SudParis, Institut Polytechnique de Paris, France \\
\email{\{firstname.last\_name\}.telecom-sudparis.eu} \and
Bonitasoft, France\\
\email{\{firstname.last-name\}@bonitasoft.com}}
\maketitle              
\begin{abstract}
Automated process discovery from event logs is a key component of process mining, allowing companies to acquire meaningful insights into their business processes. Despite significant research, present methods struggle to balance important quality dimensions: fitness, precision, generalization, and complexity, but is limited when dealing with complex loop structures. This paper introduces Bonita Miner, a novel approach to process model discovery that generates behaviorally accurate Business Process Model and Notation (BPMN) diagrams. Bonita Miner incorporates an advanced filtering mechanism for Directly Follows Graphs (DFGs) alongside innovative algorithms designed to capture concurrency, splits, and loops, effectively addressing limitations of balancing as much as possible these four metrics, either there exists a loop, which challenge in existing works. Our approach produces models that are simpler and more reflective of the behavior of real-world processes, including complex loop dynamics. Empirical evaluations using real-world event logs demonstrate that Bonita Miner outperforms existing methods in fitness, precision, and generalization, while maintaining low model complexity. 

\end{abstract}

\section{Introduction}
Process discovery~\cite{process-discovery-review} is one of the most prominent process mining techniques. It enables the automatic discovery of a process model that captures and explains the behavior recorded in event logs. The process model is typically represented in Petri nets~\cite{petri_net}, but other formats are also possible (e.g., in the standard Business Process Model and Notation - BPMN~\cite{BPMN}). A discovered process model must accurately reflect the behavior observed in or inferred from the event log. Specifically, the process model should (i) parse the traces contained within the log, (ii) parse traces not present in the log but likely to belong to the process that generated the log, and (iii) refrain from parsing traces unrelated to the process. These properties are known as fitness, generalization, and precision, respectively. Furthermore, the model should be as simple as possible, a characteristic often quantified through complexity measures~\cite{DBLP:conf/otm/BuijsDA12}.

Despite extensive research~\cite{process-discovery-review,DBLP:journals/is/WeerdtBVB12}, achieving a balance across the four key quality dimensions — fitness, precision, generalization, and complexity — remains a challenge. When applied to real-life event logs, most automated process discovery techniques, such as the Heuristics Miner~\cite{FHM} and its derivatives, often produce large, spaghetti-like models that are behaviorally incorrect (e.g., prone to deadlocks). Another state-of-the-art technique, the Inductive Miner~\cite{IM}, tends to generate block-structured models that are behaviorally sound with high fitness but exhibit low precision. Additionally, the SplitMiner approach~\cite{splitminer} is capable of producing accurate and simple process models; however, these models become complex and unsound in the presence of loops, complicating the observed behavior in the event log and reducing the accuracy and structure of the results.

\begin{figure}[h]
    \centering
    \begin{subfigure}{0.48\linewidth}
        \centering
        \includegraphics[width=\linewidth]{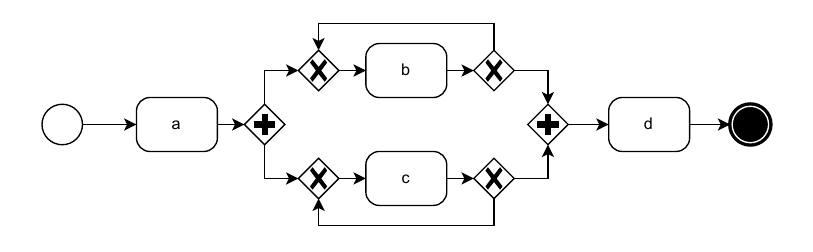}
        \caption{Discovered Model by Split Miner, Inductive Miner, Heuristics Miner.}
        \label{fig:sub:motivated_example_splitminer}
    \end{subfigure}
    \hfill
    \begin{subfigure}{0.48\linewidth}
        \centering
        \includegraphics[width=\linewidth]{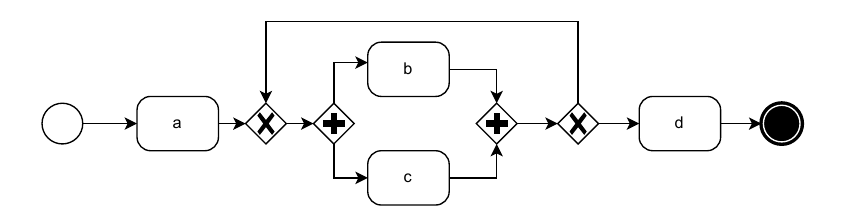}
        \caption{BPMN to Discover}
        \label{fig:sub:motivated_example_bonitaminer}
    \end{subfigure}
    \caption{A simple example of the limitation of existing works.}
    \label{fig:motivated_example}
\end{figure}

We provide a basic example to illustrate a limitation addressed by existing methods, shown in Figure~\ref{fig:motivated_example}. Consider an event log for the BPMN model represented in Figure~\ref{fig:sub:motivated_example_bonitaminer}, where there exists a \textbf{loop} that includes the \textbf{parallel} blocks \textit{b} and \textit{c}. Current state-of-the-art algorithms are unable to accurately discover this BPMN pattern with a loop over parallel blocks. Instead, they produce a model like the one shown in Figure~\ref{fig:sub:motivated_example_splitminer}, where these algorithms incorrectly represent \textit{b} and \textit{c} with individual self-loops. This misrepresentation significantly impacts the model's accuracy, reducing its usability for real-world process analysis.

To address this limitation and bridge the identified gap, this paper introduce Bonita Miner, a novel process discovery method to generate simpler process models while still achieving a balance among fitness, precision, and generalization. The approach leverages a novel Depth-First Algorithm (DFA) for constructing splits and joins, coupled with advanced filtering of the directly-follows graph (DFG) derived from event logs. These constructs effectively capture concurrency, choices, causal relationships, and even loops within the DFG. The DFA reduces model complexity while maintaining a balance among evaluation metrics, making the method suitable for processes of varying complexity, including those with loops. We also treat the loops as blocks (with one or multiple sources and one or multiple targets).

We empirically demonstrate that Bonita Miner surpasses three state-of-the-art baselines by evaluating its performance on BPIC real-life and synthetic loops event logs using eight comprehensive metrics that measure the four key quality dimensions, even in the presence of complex loop structures.

The remainder of the paper is structured as follows. Section~\ref{sec:relatedworks} reviews the existing process discovery methods. Section~\ref{sec:approachoverview} introduces the Bonita Miner method, followed by Sections \ref{sec:preliminaries} and \ref{sec:approach}, which detail the proposed approach. Section~\ref{sec:evaluation} discusses the empirical evaluation of Bonita Miner, and finally, Section~\ref{sec:conclusion} concludes the paper and outlines potential directions for future research.

\section{Related work}
\label{sec:relatedworks}
The section provides an overview of existing automated process discovery methods and discusses their limitations.

The \(\alpha\)-algorithm~\cite{DBLP:journals/tkde/AalstWM04} is a simple automated process discovery method based on the concept of DFG. While appealing due to its simplicity, the \(\alpha\)-algorithm is not applicable to real-life event logs since it assumes the log to be complete and 
is too sensitive to infrequent behavior. The Heuristics Miner~\cite{DBLP:conf/cidm/WeijtersR11} addresses these limitations and consistently performs better in terms of accuracy on incomplete and noisy logs. To handle noise, the Heuristics Miner relies on a relative frequency metric between pairs of event labels. However, while this technique demonstrates higher fitness, it faces the challenge of lower precision.

Structured process models are generally more understandable than unstructured ones~\cite{DBLP:conf/bpmn/DumasGP10,DBLP:conf/caise/DumasRMMRS12}. Moreover, structured models are sound, provided that the gateways at the entry and exit of each block match. Given these advantages, several algorithms have been designed to discover
structured process models, represented for example as process trees. 
The Inductive Miner~\cite{DBLP:conf/bpm/LeemansFA13} uses a divide-and-conquer approach to discover process structured models, achieving high fitness. However, it tends to over-generalize the behavior observed in the log whenever the process model to be discovered is unstructured, generating undesired \textit{flower models}\cite{5990012}.

The Structured Miner~\cite{DBLP:conf/er/AugustoCDRB16} addresses this limitation by relaxing the requirement of always producing a structured process model, in favor of achieving higher accuracy. Split Miner~\cite{splitminer} is another method that addresses the balance between fitness, precision, and generalization. It focuses on filtering directly-follows graphs and discovering split gateways to produce simple, accurate, and deadlock-free process models. However, Split Miner is limited in handling loops effectively, which can lead to challenges in processes with loop-intensive behavior.

To reduce these challenges, an improved approach is proposed, which enhances the handling of loops while maintaining a balance across other quality dimensions. This approach ensures better adaptability to loop-intensive scenarios, thereby improving model quality and applicability.

\section{Approach Overview}
\label{sec:approachoverview}
\begin{figure}[h]
\includegraphics[width=\linewidth]{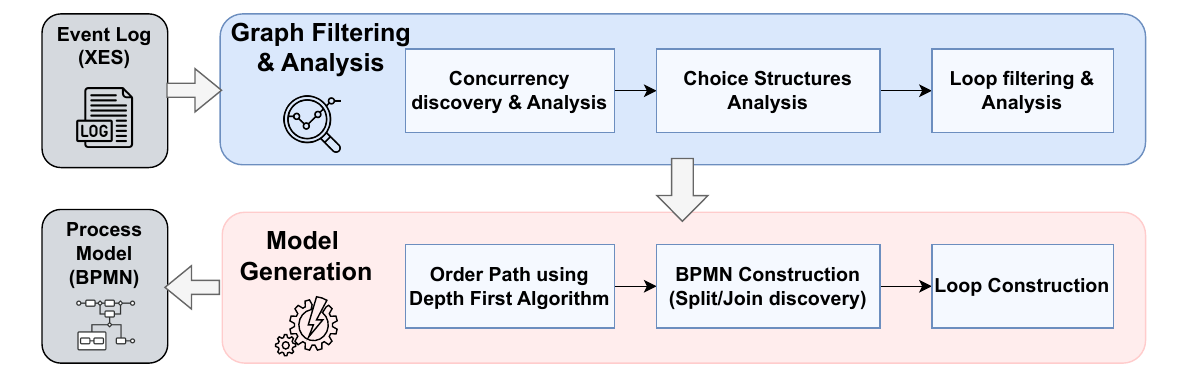}
\centering
\caption{Bonita Miner Overview}
\label{fig:overview}
\end{figure}

As mentioned in the introduction, the goal of the algorithm introduced in this paper is to ensure the balancing of key metrics even in the presence of loops. For example, it successfully identifies the pattern represented in the BPMN model of Figure~\ref{fig:sub:motivated_example_bonitaminer}, achieving a model accuracy of 1.
The proposed approach decomposes the process discovery task in two main steps, as shown in Figure~\ref{fig:overview}. 

\medbreak

\textbf{Graph filtering and analysis:} taking an event log as input, this step first transforms it into a graph. Then, by analyzing both the graph and the event log, concurrent activities, choices, and part of the loops are removed from the graph. The benefits of processing these behaviors separately from the graph is twofold: it enables (i) handling complex loops (e.g., from one or multiple sources to one or multiple targets) and (ii) generating the BPMN model with a single-shot approach. with existing methods that require multiple iterations over the process model to achieve correct semantics, particularly for handling complex behaviors such as nested splits and joins.


\textbf{Model Generation:} after filtering the graph to remove loops and concurrency, the next step involves ordering all paths in the graph using a depth-first algorithm. This approach ensures that joins are introduced alongside splits, constructing blocks of split/join gateways without enforcing strict block-structuredness. The resulting model is a BPMN diagram\footnote{This approach is not limited to BPMN; the same steps can also be applied to generate a Petri net.}, where elements, including gateways (and nested gateways), are iteratively added based on insights from the earlier analysis. This step enables the construction of the BPMN diagram without loops in a single iteration, eliminating the need for post-processing or replacing gateways to achieve correct model semantics. Loops are constructed separately, employing a strategy that reduces complexity while maintaining a balance in quality metrics. This approach contrasts with existing methods, which typically handle loops alongside other tasks, resulting in increased complexity.


These steps enable the construction of a simplified model with low complexity, while achieving a high balance of fitness, precision, and generalization.

\section{Preliminary}
\label{sec:preliminaries}
In this section, we give some definitions related to event logs and process models that will be used in the subsequent sections. 

A process model is a set of connected entities that formally represent process behavior. Since we use BPMN as the modeling language, entities can include flow objects (activities, events, and gateways), swimlanes (pools and lanes), artifacts (e.g., data objects, text annotations, or groups), and connecting objects (sequence flows, message flows, and associations). In this work, we limit the scope to entities shared among different languages, specifically flow objects represented by \emph{activities}, \emph{start/end events}, and \emph{gateways} (including parallel and exclusive gateways), and connecting objects represented by \emph{sequence flows}. 

\begin{definition}[Event Log] Given a set of events $\mathcal{E}$, an \textit{event log} $\mathcal{L}$ is a multiset of traces as $\mathcal{T}$, where a trace $t \in \mathcal{T}$ is a sequence of events $t = \langle e_1, e_2, \dots, e_n \rangle$, with $e_i \in \mathcal{E}$, $1 \leq i \leq n$. Additionally, each event has a label $l \in \mathcal{L}$ and it refers to a task/activity executed within a process, we retrieve the label of an event with the function $\lambda: \mathcal{E} \to \mathcal{L}$, using the notation $\lambda(e) = e^l$.
\end{definition}
For the remaining, we assume all the traces of an event log have the same start event and the same end event. This is guaranteed by a simple preprocessing of the event log, to be compliant with the third of the 7PMG in ~\cite{7PMG}, i.e., ``use one start event for each trigger and one end event for each outcome''.

Given the set of labels $\mathcal{L} = \{s, a, b, c, d, e, f, h, i, j, k, l, end\}$, a possible log is:
$
\mathcal{L} =
\{
\langle s, a, b, end \rangle^{10},
\langle s, a, c, e, f, h, d ,i, j, k, l, end \rangle^{10},
\langle s, a, c, e, f, h, d ,j, i, k, l, end \rangle^{10},$ \\$ 
\langle s, a, c, f, e, h, d ,i, j, k, l, end \rangle^{10},  \dots
\};$
this log contains more than 10 distinct traces, each of them recorded 10 times.

Starting from a log, we construct a DFG in which each arc is annotated with a frequency, based on the following definitions.



\begin{definition}[Directly-Follows Graph]
\label{def:dependency_graph} Given an event log $\mathcal{L}$, its directly follows graph is defined as a directed graph $DFG = (V, E)$, where:
\begin{itemize}
    \item $V$ represents a finite set of vertices, each corresponding to an activity or start/end event, and having a unique label $l \in \mathcal{L}$.
    \item $E \subseteq V \times V$ represents a set of directed edges, where each edge $(u, v) \in E$ signifies that $v$ can be executed right after the completion of $u$, theoretically $\forall (\lambda(a)=u$ and $\lambda(b)=v), \exists t= \langle e_1,e_2,\dots,e_n \rangle \in \mathcal{T}$ where $a = e_i$ and $b = e_{i+1}$.
   \item $v_s \in V$ is the starting vertex, iff $ \nexists (v, v_s) \in E$.
   \item $v_e \in V$ is an ending vertex, iff $\nexists (v_e,v) \in E$.
    \item For $e = (u,v) \in E$, we use the notations $e.source = u$ and $e.target = v$. 
   \item For $v \in V$, the direct successors of $v$ is $\operatorname{Succ}_{DFG}(v) = \{ u \in V \mid (v,u) \in E\}$ and the direct predecessors of is $\operatorname{Pred}_{DFG}(v) = \{ u \in V \mid (u,v) \in S\}$. When it is clear from context, the subscript $DFG$ is omitted.
   \item For $(u,v) \in E$, we denote $\|u \rightarrow v \|$ = $\sum |(u,v)|$ that represent the frequency of the edge from $u$ to $v$, where $\forall (\lambda(a)=u$ and $\lambda(b)=v), \exists t = \langle e_1,e_2,\dots,e_n \rangle \in \mathcal{T}$ where $a = e_i$ and $b = e_{i+1}$. 
   \item For $(u,v) \in E$, we denote by short loop $u\circlearrowleft v$ where $\exists (\lambda(a)=u$ and $\lambda(b)=v), \exists t = \langle e_1,e_2,\dots,a,b,\dots,b,a,\dots ,e_n \rangle \in \mathcal{T}$.  

\end{itemize}
\end{definition}

Let $DFG = (V, E)$ be a DFG. We define paths, cycles and branching relations are as follows:
\begin{definition}[Path]
A path from $u \in V$ to $v \in V$, denoted as $P_{u,v} = \langle (v_1,v_2), \dots, (v_{n-1}, v_n) \rangle$, is the sequence of unique edges leading from $u$ to $v$ where $\forall 1 \leq i,j \leq n-1, i \neq j, n \geq 2, ~(v_i, v_{i+1}) \in E$, $v_1 = u, v_n = v$, and $\nexists (v_i, v_{i+1}) = (v_j, v_{j+1})$. We denote by:
\begin{itemize}
    \item $P^v_{u,v} = \{v_1, \dots, v_n\}$: the set of the path vertices. 
    \item $P^s_{u,v} = (v_1, v_2)$: the first edge of the path.
    \item $P^e_{u,v} = (v_{n-1},v_n)$: the last edge of the path.  
    \item $\mathbb{P}_{u,v}$: the possibly empty set of all paths from $u$ to $v$.
\end{itemize}
\end{definition}

\begin{definition}[Cycle]
\label{def:cycle}
A cycle $C_u$ is a path $P_{u,u}=\langle (u,v_1), \dots, (v_n, u)\rangle$ that starts and ends with $u$ such that $\forall 1 \leq i \leq n, v_i \neq u$. We denote by $\mathbb{C}_u $ the possibly empty set of all cycles of $u$.
\end{definition}



\begin{definition}[BPMN Process Model]
\label{def:dependency_graph}
A BPMN process model is defined as $M = (N, G, S)$, where:
\begin{itemize}
    \item $N = V$ is the set of vertices in $DFG$.
    \item $G$ is a finite set of gateways. A gateway $g \in G$ can be parallel, represented by $+$ or exclusive represented by $\times$. We denote by $type(g) \in \{\times, +\}$ the function that returns the type of g.
    \item $S \subseteq (N \times G) \cup (N \times G) $ is a set of directed sequence flows. For each $s = (u,v) \in S$, we use the notations s.source = u, and s.target = v. 
     \item For $u \in N \cup G$, the direct successors of u is $\operatorname{Succ}_M(u) = \{ v \in N \cup G \mid (u, v) \in S\}$ and the direct predecessors is $\operatorname{Pred}_M(u) = \{ v \in N \cup G \mid (v,u) \in S\}$. When it is clear from context, subscript $M$ is omitted.
     \item For $u \in N \cup G$, the transitive successors of $u$ is $\operatorname{TSucc}(u) = \{v \in N \mid \exists g_1, \ldots, g_k \in G, (u, g_1), (g_k, v), (g_i, g_{i+1}) \in S\}$ and the transitive predecessors of $u$ is $\operatorname{TPred}(u) = \{v \in N \mid \exists g_1, \ldots, g_k \in G, (v, g_1), (g_k, u), (g_i, g_{i+1}) \in S\}$. For $u \in N$, $\operatorname{TSucc}(u) = \operatorname{Succ}_{G}(u)$ and $\operatorname{TPred}(u) = \operatorname{Pred}_{G}(u)$.
\end{itemize}
\end{definition}

\section{Bonita Miner}
\label{sec:approach}
In this section, we illustrate the details of our approach. We begin by running the example in Section~\ref{subsec:running_example}, followed by filtering the concurrent activities in Section~\ref{subsec:concurrency}. Next, we identify the loops in Section~\ref{subsec:looping_edge}. We then delve deeper into identifying the activities that represent transitivity after the gateways, including the identification of loop blocks in Section~\ref{subsec:synthesisFlow}. This is followed by the construction of split and join patterns in the model in Section~\ref{subsec:split_join}, and finally, the loop construction in the BPMN in Section~\ref{subsec:loop_construction}.

\subsection{Running Example}
\label{subsec:running_example}

\begin{figure}[h]
\includegraphics[width=\linewidth]{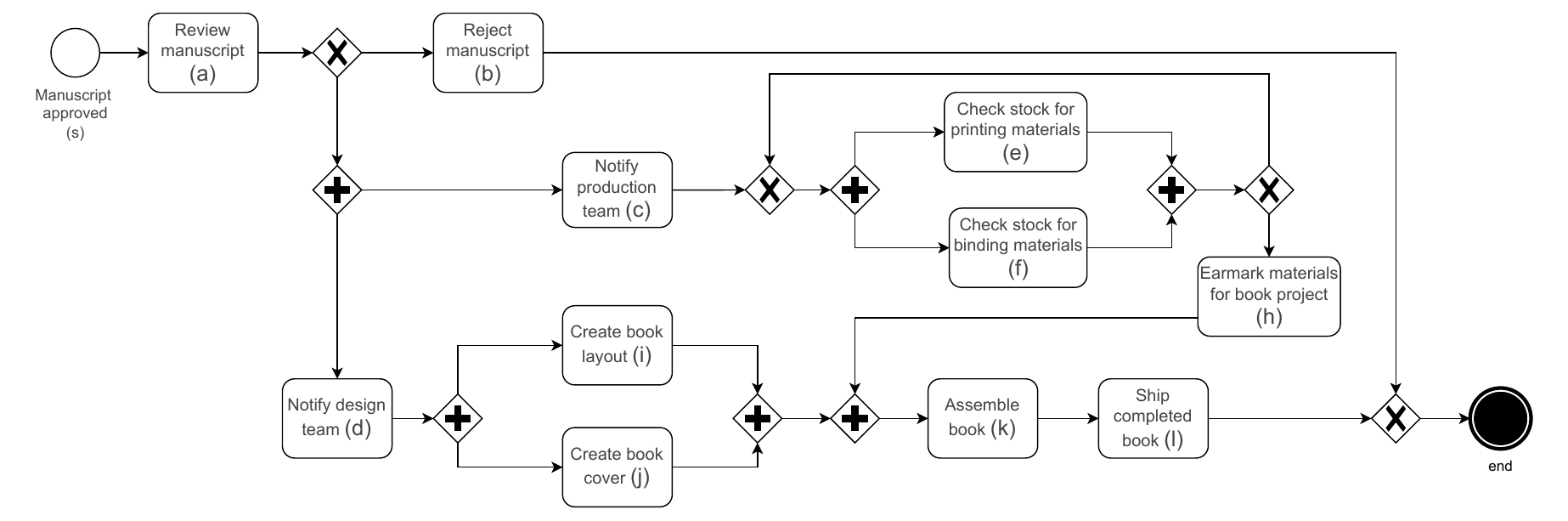}
\centering
\caption{The running example is used to explain the Bonita Miner workflow}
\label{fig:runing_example}
\end{figure}

Figure~\ref{fig:runing_example} represents a BPMN process model. After activity (a), there are nested gateways: an XOR gateway connected to activity (b) and an AND gateway that connects to two activities, (c) and (d). Additionally, there are two nested join gateways before activity (h). Moreover, there is a loop on a parallel block encompassing activities (e) and (f). Since these patterns are present in the example, this model serves as a running example to illustrate the steps of the algorithm.

\subsection{Concurrency discovery}
\label{subsec:concurrency}
\begin{figure}[h]
\includegraphics[width=\linewidth]{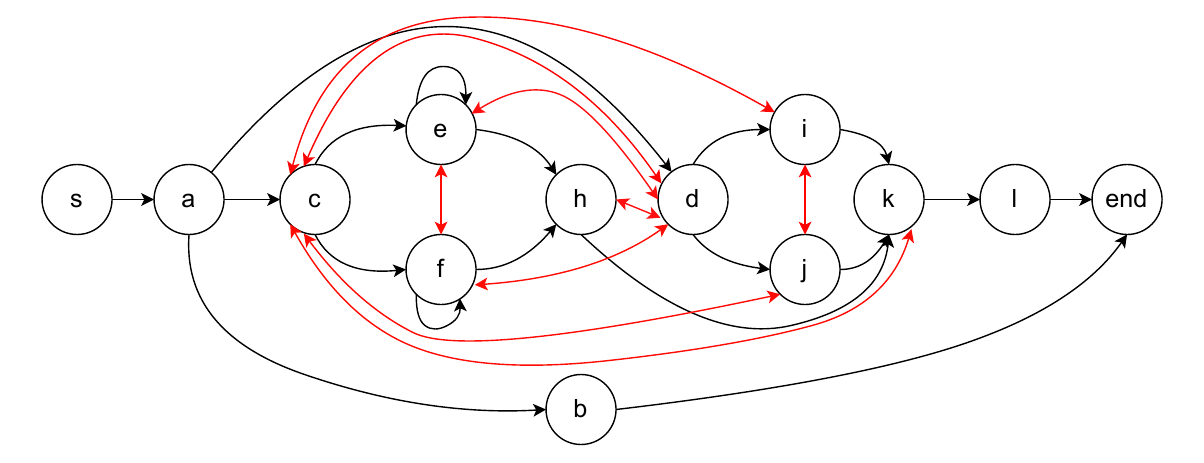}
\centering
\caption{Part of the DFG of the event log of the running example}
\label{fig:dfg}
\end{figure}

Identifying concurrency and filtering it effectively is crucial for simplifying the DFG. With a simplified DFG, we can extract branching relations and the dependent relations between activities. Additionally, concurrency discovery helps construct AND gateways (including both simple and nested gateways) during one-step generation.
An event log is the input of our algorithm, and it can be converted into a DFG (Definition~\ref{def:dependency_graph}), as illustrated in Figure~\ref{fig:dfg}. To enhance the readability of the DFG, we add two arrows on each relation to represent the existence of two connections between a pair of activities. For example, $c \leftrightarrow d$ indicates that there exists a relation $c \rightarrow d$ as well as a relation $d \rightarrow c$. For the sake of clarity, we represent only a subset of these relations (we remove the parallelism that arises from the activities (e), (f), (h), (i), (j), and (k)). All the red relations in the DFG represent parallelism. The first step is to detect all concurrent relations, formally defined in Definition~\ref{def:parallel_relation}.

\begin{definition}[Concurrent Relation]
\label{def:parallel_relation}
Let an event log $\mathcal{L}$, a set of traces $\mathcal{T}$, and a directed graph $\mathrm{DFG} = (V, E)$ be given. We postulate that $a$ and $b$ are concurrent if and only if the following two conditions hold:
\begin{enumerate}
    \item $\exists e, e' \in E$ such that $ a, b \in V$ and $(a, b) = e$ and $(b, a) = e'$ 
    \item $\exists t \in \mathcal{T}$ such that $\exists a,b \in t$
\end{enumerate}
\end{definition}

After identifying all concurrent relations, we remove them except for those representing short loops, as we are focusing on parallelism in this step. This step is very important for filtering the graph and subsequently identifying the gateway type in the next step. In the following definitions and sections, we use the term DFG to refer to a filtered graph that excludes concurrency relations (see Figure~\ref{fig:dependency_graph}).

Branching parallel relations are those derived from concurrency relations that share the same predecessor. This characteristic helps in detecting parallel split gateways (AND gateways), particularly by identifying the activities that will transitively follow a gateway. For example, $\{c, d\}$ in the filtered DFG shares the same predecessor $(a)$; hence, it is added to the set of branching parallel relations. However, other relations, such as $\{c, e\}$, are not included because they do not have the same predecessor. 
 As a result, the branching parallel relations are: $\{\{c, d\}, \{i, j\}, \{e, f\}\}$.

Formally, the branching parallel relations is defined as follow:
\begin{definition}[Branching Parallel Relations]
\label{def:parallelism}
Let $\mathrm{DFG}' = (V', E')$ be the filtered DFG, where concurrency relations have been removed except for short loops. 
The set of branching parallel relations is denoted as:
\[
BP = \{\{a, b\} \mid a, b \text{ are concurrent and } \operatorname{Pred_{DFG'}}(a) = \operatorname{Pred_{DFG'}}(b)\}.
\]
\end{definition}

\subsection{Loop filtering}
\label{subsec:looping_edge}
\begin{figure}[h]
\includegraphics[width=0.8\linewidth]{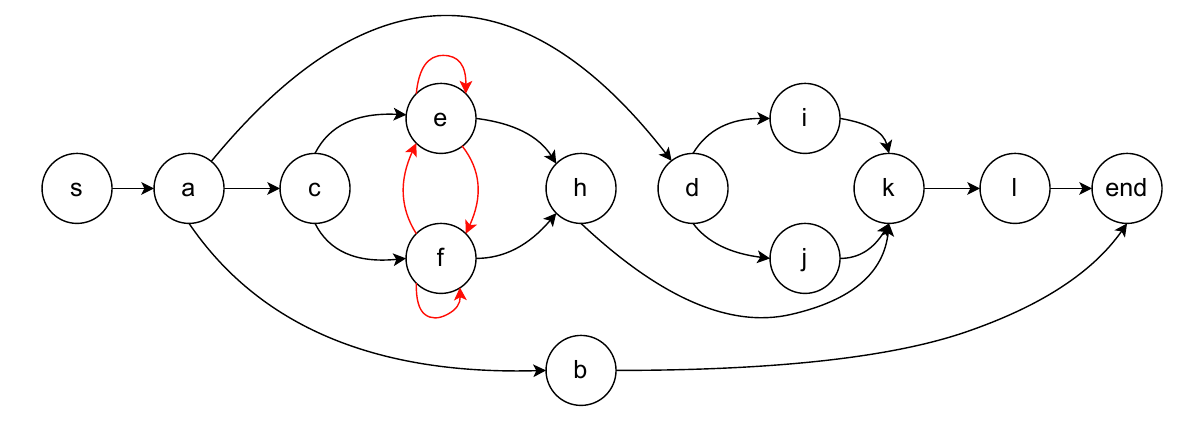}
\centering
\caption{DFG, after filtering the parallelism, of the running example with cycles}
\label{fig:dependency_graph}
\end{figure}

Although loops contribute to increased complexity and reduced readability in a process model, we treat loops in our approach as blocks, which may have one or multiple sources and one or multiple targets. This treatment and construction of loops as blocks help to reduce complexity significantly. This step involves removing the loops from the $DFG$ by detecting and filtering the dependency relations in $E$ that break the cycles which results in an acyclic DFG. For example, given the DFG in Figure~\ref{fig:dependency_graph}, there are four cycles ${C_{e}}_1 = \langle (e,e) \rangle$, ${C_{e}}_2= \langle (e,f), (f,e) \rangle$, ${C_{f}}_1 = \langle (f,f) \rangle$ and ${C_{f}}_2 = \langle (f,e), (e,f) \rangle$. 

Breaking cycles involves removing specific edges, known as \emph{looping edges}, which are the last edges in the paths that create the looping behavior. For example, given the cycle ${C_{e}}_2 = \langle (e,f), (f,e) \rangle$, the edge $(f,e)$ creates the loop. It is characterized by the property that, following the path from the graph's start to the edge's source vertex, no other vertices in the cycle are revisited before reaching this edge. In this example, the looping edges are $(e,e)$, $(f,e)$, $(f,f)$, and $(e,f)$. Thus, all the looping edges are removed to break the loops.

\begin{figure}[h]
\includegraphics[width=0.5\linewidth]{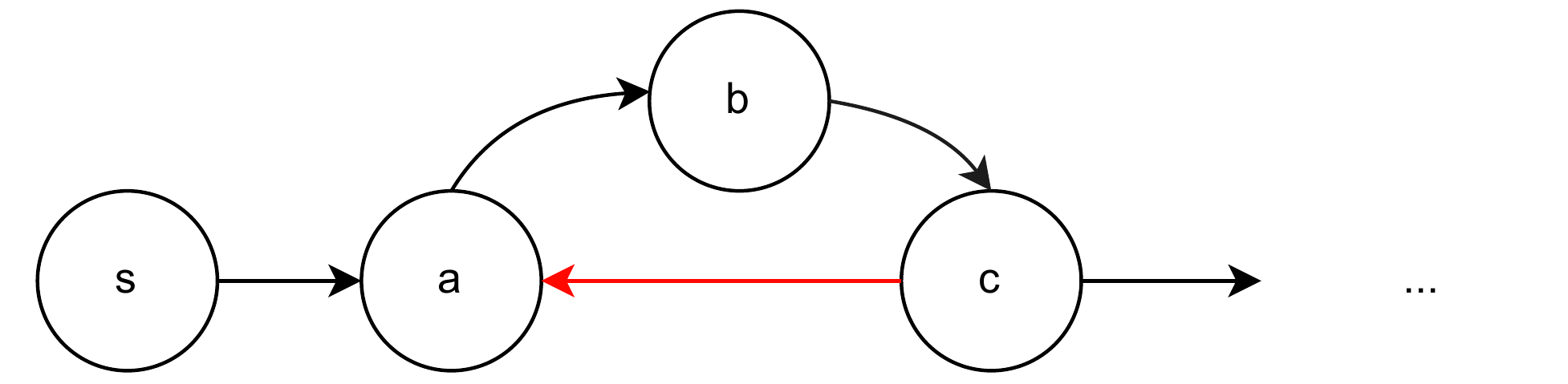}
\centering
\caption{An example of a long loop.}
\label{fig:dependency_graph_long_loop}
\end{figure}

The DFG in Figure~\ref{fig:dependency_graph_long_loop} provides another example. The cycles in this example are $C^1_{a}=\langle(a,b), (b,c), (c,a)\rangle$, $C^1_{c}=\langle(c,a), (a,b), (b,c)\rangle$, and $C^1_{b}=\langle(b,c), (c,a), (a,b)\rangle$. To break the loop, following the path from the graph's start to the edge's source vertex of the cycles, the looping edge is characterized by the property that no other vertices in the cycle are revisited before reaching this edge. Applying the path condition in Definition~\ref{def:looping_edge}, only $C^1_{a}$ is valid. Therefore, $L=\{(c,a)\}$.

These are formalized in Definition~\ref{def:looping_edge} as follows:

\begin{definition}[Looping edge]
\label{def:looping_edge}
Let $C_u$ be a cycle of $u \in V$. The looping edge \(e^{\text{loop}}\) is an edge \(e \in C_u^E\) such that $P^V_{v_s, C_u^S.\text{source}} \cap \left(C_u^V \setminus \{C_u^S.\text{source}\} \right) = \emptyset$.
\end{definition}


After filtering all the looping edges, an acyclic DFG, denoted as $DFG_A$, is obtained and used in the next step to discover the split and join gateways. The result after extract looping edge is the removal of the red edges from the Graph in the Figures~\ref{fig:dependency_graph} and \ref{fig:dependency_graph_long_loop}.

\subsection{Synthesis of Control Flow Elements}
\label{subsec:synthesisFlow}
In the following subsections, we illustrate the analysis of choices, concurrency, and loops to develop a specific format that aids in the construction phase of the BPMN. This format is designed to detect gateway types, especially in the presence of nested gateways, and to preprocess loops by constructing blocks before adding them to the BPMN.

\subsubsection{Exclusive \& Concurrency Relations Analysis}
In this step, we extract all the choices between the activities, where the choices can be represented as branching without the existence of parallelism (Definition ~\ref{def:parallelism}).
For example, in Figure~\ref{fig:dfg}, after (a), there are three activities: (b), (c), and (d). Since (c) and (d) are in parallel, we can deduce that (b) and (c), as well as (b) and (d), are in choice due to the transitivity of an XOR gateway. These are formalized in Definition~\ref{def:choice} as follows:

\begin{definition}[Branching Exclusive Relations]
\label{def:choice}
Let $BP$ be the set of Branching Parallel relation. The $BE_u$ is a set of exclusive branching after $u \in V$, where $\forall \{a,b\} \in BE_u, \exists (u,a) \in E$ and $(u,b) \in E$ where $a \neq b$ and $\{a,b\} \notin BP$. 
\end{definition}

We denote $BE = \bigcup_{v \in V} BE_v$ to represent all the branching choice in the $DFG_A$.

After extracting the branching exclusive relations, we analyze these relations to explore transitivity after a gateway between the activities and merge certain branching relations. This helps our algorithm detect the nesting gateways in our main algorithms. The following Algorithm~\ref{algo:merge_exclusive} represents the process of merging these relations. Before starting, we arrange the subsets of $BE$ into a list based on the most frequent vertices in the subsets. This step is crucial to account for all possible relations, especially in cases involving multiple nested gateways (e.g., an XOR gateway followed by an AND gateway, which is then followed by another XOR gateway). Starting with the most frequent vertices ensures that all gateways are detected effectively.

\begin{algorithm}[h]
\footnotesize
\begin{algorithmic}[1]
\STATE \textbf{Input:} Acyclic DFG $DFG_A$, Branching Exclusive Relation $BE$
\STATE \textbf{Output:} Merged Exclusive denoted by $\#$
\STATE create a empty set $\#$
\FORALL{$b \in BE$}
    \FORALL{$b' \in BE \setminus \{b_1\}$}
            \IF{$b \cap b' \neq \emptyset$}
                 \IF{$Succ(b.source) \cap Succ(b.target) \cap Succ(b'.source) \cap Succ(b'.target) \neq \emptyset$}
                \STATE $\# \gets \# \cup \{ \{b.source, b.target, b'.source, b'.target \}\}$
            \ENDIF
        \ENDIF
    \ENDFOR
    \STATE $BE \gets BE \setminus \{b\}$ 
\ENDFOR

\end{algorithmic}
\caption{Merge Exclusive}
\label{algo:merge_exclusive}
\end{algorithm}

For example, from the $DFG_A$ in Figure~\ref{fig:dependency_graph}, $BE = \{\{b, c\}, \{b, d\}\}$. Since \(b\) is the command vertex (line 6) and all the vertices \(b\), \(c\), and \(d\) share the same successor \(a\) (line 7), the result is \(\# = \{\{b, c, d\}\}\). We denote $\#(v_1,v_2)$ is the smaller subset that contains $v_1 \in V$ and $v_2 \in V$.

Similar to the Merge Exclusive (Algorithm~\ref{algo:merge_exclusive}), the Merge Parallel is based on the Branching Parallel Relations (Definition~\ref{def:parallelism}). The same algorithm is then applied to merge the parallel relations, and we denote the result by \(\|\). After applying the algorithm, \(\| = \{\{c, d\}, \{i, j\}, \{e, f\}\}\). We denote $\|(v_1,v_2)$ is the smaller subset that contains $v_1 \in V$ and $v_2 \in V$, and $\|(v)$ to represent all the subset that contain $v$.



\subsubsection{Loop Structures Analysis}
After extracting loops (as described in Definition~\ref{def:looping_edge}), we preprocess the loops to form blocks of loops, which helps reduce unnecessary gateways and relations used to represent the loops. To achieve this, two merge algorithms are applied to combine the loops based on their sources and targets. These algorithms assist in identifying the start and end points of the block where the loop should be constructed.

The "Merge Loop by sources" algorithm~\ref{algo:merge_loop_source} is designed to merge loops based on an acyclic dependency graph based on their structure and relationships, and based on the looping edge defined in ~\ref{def:looping_edge}. This merge groups loops that share the same source and whose targets have common successors (line 7-9). In the example, the looping edge $L = \{(f,f), (e,e), (f,e), (e,f)\}$, and the result from this algorithm is $ML_s = \{(f,\{f,e\}), (e,\{e,f\})\}$, which is the input of the second algorithm~\ref{algo:merge_loop_target}.

\begin{algorithm}[h]
\footnotesize
\begin{algorithmic}[1]
\STATE \textbf{Input:} Acyclic DFG $DFG_A$, looping edge $L$
\STATE \textbf{Output:} Merged loops $ML_s$
\STATE create a list $ML_s$
\FORALL{$l \in L$}
    \STATE create a set $targets \gets \{l.target\}$
    \FORALL{$l' \in L \setminus \{l\}$}
        \IF{$l.source == l'.source ~\wedge~ \operatorname{Succ}_{DFG_A}(l.target) == \operatorname{Succ}_{DFG_A}(l'.target)$}
            \STATE $targets \gets targets \cup \{ l'.target \}$
            \STATE $L \gets L \setminus \{l'\}$ 
        \ENDIF
    \ENDFOR
    \STATE  $ML_s \gets ML_s \cup (l.source,targets)$
    \STATE $L \gets L \setminus \{l\}$ 
\ENDFOR

\end{algorithmic}
\caption{Merge Loop by sources}
\label{algo:merge_loop_source}
\end{algorithm}

The "Merge Loop" by targets algorithm~\ref{algo:merge_loop_target} is designed to consolidate loops that have already been grouped based on their sources. The goal is to further merge these loops by examining their targets and associated predecessors. This merge \( ML \), is derived from \( ML_s \) by combining sources if they have common predecessors (lines 7-9). In the example, the extracted $ML_s = \{(f,\{f,e\}),(e,\{e,f\})\}$, and the result from this algorithm is $ML = \{(\{f,e\},\{f,e\})\}$. This indicates the existence of a loop for the entire block of $\{e,f\}$.

\begin{algorithm}[h]
\footnotesize
\begin{algorithmic}[1]
\STATE \textbf{Input:} Acyclic DFG $DFG_A$, Loop merged by sources $ML_s$
\STATE \textbf{Output:} Merged loops $ML$
\STATE create a list $ML$
\FORALL{$l \in ML$}
    \STATE create a set $sources \gets \{l.source\}$
    \FORALL{$l' \in ML_s \setminus \{l\}$}
            \IF{$l.target == l'.target ~\wedge~ Pred_{DFG_A}(l.source) == Pred_{DFG_A}(l'.source)$}
                \STATE $sources \gets sources \cup \{ l'.source \}$
                \STATE $ML_s \gets ML_s \setminus \{l'\}$ 
        \ENDIF
    \ENDFOR
    \STATE  $ML \gets ML \cup (sources,l.target)$
    \STATE $ML_s \gets ML_s \setminus \{l\}$ 
\ENDFOR

\end{algorithmic}
\caption{Merge Loop}
\label{algo:merge_loop_target}
\end{algorithm}

\subsection{Split \& Join Discovery}
\label{subsec:split_join}
Accurately discovering split and join gateways while keeping the model simple is challenging. To simplify model generation, we construct blocks of split/join gateways without enforcing the block-structuredness property. Algorithm~\ref{algo:bpmn_construction} outlines the high-level steps. It takes as input the acyclic DFG $DFG_A$ and the branching parallel relations $\|$ and returns as output a BPMN model $M$.

Initially, we traverse all paths in \( DFG_A \) from start to all possible ends, ordering them with a depth-first algorithm based on the longest path. This ensures joins are introduced alongside splits, focusing on identifying the maximum number of gateways by analyzing the longest paths first. The result is a list of ordered edges derived from these paths (Line 5). In our example of Fig.~\ref{fig:dependency_graph} (excluding the red looping edges that have been filtered), the ordered depth-first traversal returns the paths ordered $[P_1, P_2, P_3, P_4, P_5]$ where $P_1=\langle(s,a), (a,c), (c,e), (e,h), (h,k), (k,l), (l, end)\rangle $, $P_2=\langle(s,a), (a,c), (c,f), (f,h),$ $(h,k), (k,l), (l, end)\rangle $, $P_3=\langle(s,a), (a,d), (d,i),$ $ (i,k), (k,l), (l, end)\rangle $, $P_4=\langle(s,a),$ $(a,d), (d,j), (j,k), (k,l), (l, end)\rangle $, and $P_5$ $=\langle(s,a), (a,b), (b, end)\rangle$. The ordered list of unique edges $E_O$ extracted from these paths, respecting their order, is $E_O=[(s,a), (a,c), \dots, (c,f), (f,h) \dots]$.

\begin{algorithm}[h] 
\footnotesize
\begin{algorithmic}[1]
\STATE \textbf{Input:} Acyclic DFG $DFG_A=(V,E)$, parallel vertices $\|$, exclusive vertices $\#$
\STATE \textbf{Output:} Generated BPMN Model $M=(N,G,S)$
\STATE Initialize $M$
\STATE Create a gateway reference $\textit{GR} = null$
\STATE $E_O = \operatorname{OrderedDF}(DFG_A)$
\FORALL{$e \in E_O$}
    \STATE $source \gets e.source$
    \STATE $target \gets e.target$
    \IF{$\nexists s', s'' \in S ~|~ s'.source=source ~\wedge~ s''.target = target$}
        \STATE $N \gets N \cup \{source, target\}$
        \STATE $S \gets S \cup \{(source,target)\}$
     \ELSIF{$\exists s \in S ~|~s.source = source ~\wedge~ target \notin N$}
             \STATE call $AddSplitGateways(DFG_A, M, \|, \#, source,target, GR)$
    \ELSE
           \STATE call $AddJoinGateways(M, source, target, GR)$
    \ENDIF    
\ENDFOR
\end{algorithmic}
\caption{BPMN Construction}
\label{algo:bpmn_construction}
\end{algorithm}

All the edges of $P_1$ are added in temporal order in the process model. Lines 9-11 add the source and target elements that are not already added to $M$. From $P_2$, the dependency $(c, f)$ should be added, where $c$ already exists and has a relation. Therefore, the split algorithm is called according to line 13 (detailed in Section~\ref{subsubsection:split}, Figure~\ref{subfig:split_1}). This is followed by $(f, h)$, where the join algorithm is invoked as specified in line 15 (detailed in Section~\ref{subsubsection:join}, Figure~\ref{subfig:join_1}). From path $P_3$, for $(a, d)$, the split algorithm is called (Figure~\ref{subfig:split_2}). This is followed by $(d, i)$, which is added in line 13. Then, the join algorithm is called to add the relation between $(i, k)$. For $P_4$, $(d, j)$ is handled by calling the split algorithm, followed by the join algorithm for $(j, k)$ (Figure~\ref{subfig:join_2}). Finally, from $P_5$, $(a, b)$ is added by calling the split algorithm, and $(b, end)$ is added directly (lines 9-11).

\subsubsection{Split Operation}
\label{subsubsection:split}
\begin{figure}[h]
\begin{subfigure}{.5\textwidth}
    \centering
    \includegraphics[width=0.9\linewidth]{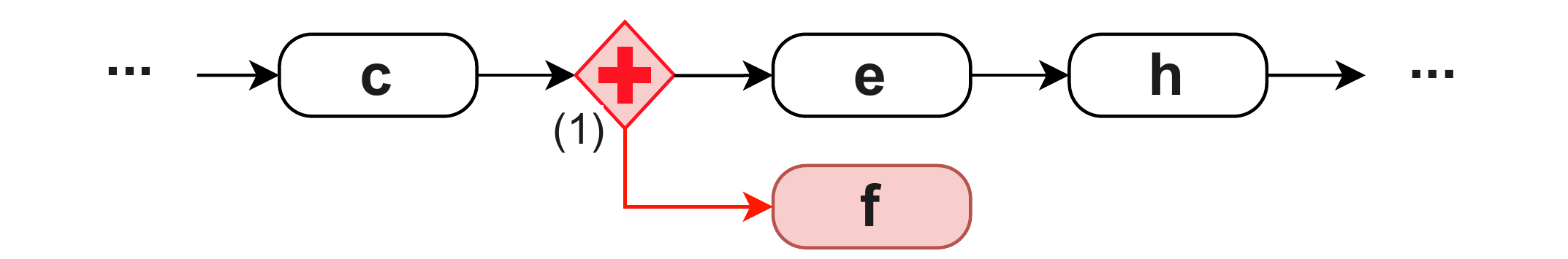}
    \caption{a simple split}
    \label{subfig:split_1}
\end{subfigure}
\begin{subfigure}{.5\textwidth}
    \centering
    \includegraphics[width=0.9\linewidth]{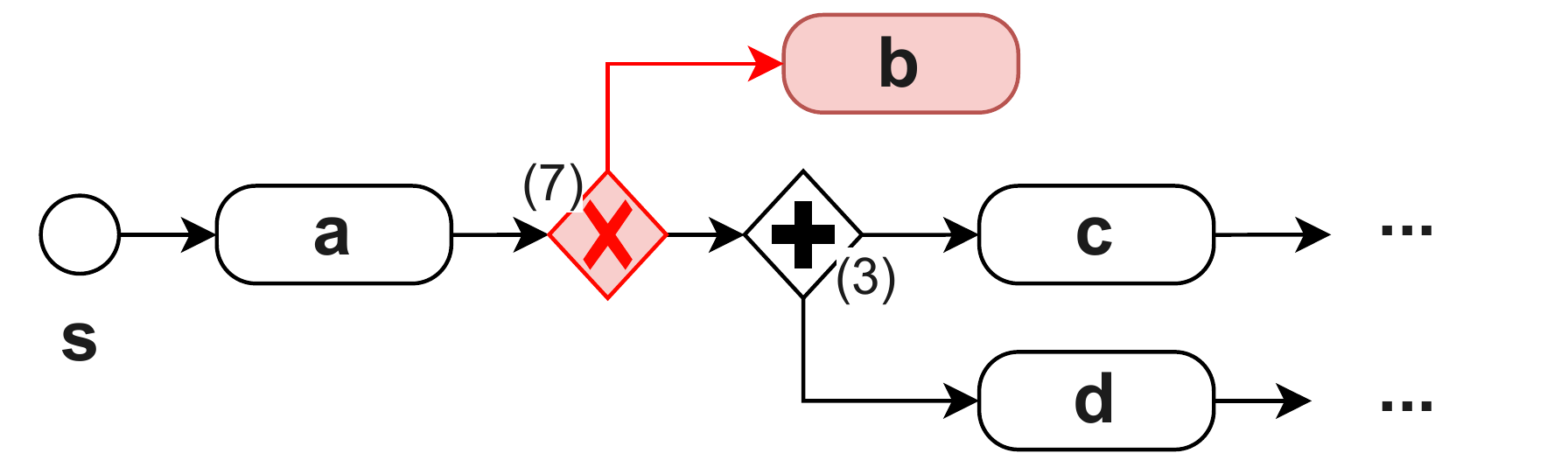}
    \caption{a nested split}
    \label{subfig:split_2}
\end{subfigure}
 \caption{Add split gateways to BPMN.}
 \label{fig:split}
\end{figure}

The split operation in Algorithm~\ref{algo:split} has two parts. The first adds a simple gateway (Figure~\ref{subfig:split_1}). The second, a recursive operation, nests gateways (Figure~\ref{subfig:split_2}).

Figure~\ref{subfig:split_1} illustrates the addition of a new simple gateway with the dependency relation $(c, f)$. Since $c$ is an existing source with a relation in the process model ($e$), the split algorithm is applied. The first step involves checking the successor of $c$, which is $e$ (lines 3-7). If the successor is an activity, a split gateway is added, with $c$ as the incoming connection and $e$ and $f$ as the outgoing connections. To determine the gateway type, the relation between $e$ and $f$ is checked. This relation must exist within the smaller subset of $\#$ and $\|$. Based on the results, if the relation is in $\|$, the gateway will be $+$; otherwise, it will be $\times$. If the successor is a gateway, the advanced split algorithm is called to add a nested gateway (lines 9-10).

\begin{algorithm}[h]  
\footnotesize
\begin{algorithmic}[1]                
\STATE \textbf{Input:} Acyclic DFG $DFG_A$, Process Model $M$, parallel vertices $\|$, exclusive vertices $\#$, Entity $source$, Entity $target$, gateway reference $GR$
\STATE $target_M \gets s.target ~|~ \exists s = (source, target_M) \in S$
\IF{$type(target_M) \notin G$}
    \STATE $g_{new} \gets Gateway(``+")$ if $\exists \{target_M,target\} \in \| ~\wedge~ (\#(target_M,target) = \emptyset$ or $\|\|(target_M,target)\| < \|\#(target_M,target)\|)$ else $Gateway(``\times")$
    \STATE $G \gets G \cup \{g_{new}\}$
    \STATE $S \gets S \setminus \{(source, g_{current}\}$
    \STATE $S \gets S \cup \{(source, g_{new}), (g_{new}, target), (g_{new}, target_M)\}$
\ELSE
    \STATE $s \gets (source, target_M)$
    \STATE $AddNestingSplitGateway(DFG_A, M, \|, \#, GR, s, source, target)$
\ENDIF
\end{algorithmic}
\caption{AddSplitGateways}
\label{algo:split}
\end{algorithm}

The advanced algorithm in Algorithm\ref{algo:split_recursive} is utilized to add a split gateway after the source, which can result in nested gateways. The algorithm involves three main steps: (i) adding the target as a successor to the existing gateway (lines 4-6), (ii) inserting a new split gateway before the existing gateway (line 7-18), and (iii) adding a new gateway after the existing one using the nesting split algorithm (lines 20-24).

For example, in Figure~\ref{subfig:split_2}, the dependency relation is $(a, b)$. The recursive algorithm begins by identifying the successor of $a$, which is the gateway $(1)$. Next, all successors with the transition $(3)$ are searched, resulting in the extracted list $\{c, d\}$. The relationship between the target $b$ and the extracted list $\{c, d\}$ is then analyzed. Since there is no parallelism in $\|$ that includes $(b, c)$ or $(b, d)$, it is concluded that $b$ is in an exclusive decision with these activities: $(b, c) \Rightarrow$ exclusive; $(b, d) \Rightarrow$ exclusive. Because this relation is exclusive and not of the same type as $(3)$, a new gateway $(7)$ is added before the existing gateway (lines 13-18).

\begin{algorithm}[H]
\footnotesize
\begin{algorithmic}[1]                
\STATE \textbf{Input:} Acyclic DFG $DFG_A$, Process Model $M$, parallel vertices $\|$, exclusive vertices $\#$, gateway reference $GR$, gateway type $gatewayType$, current relation $s$, Entity source $source$, Entity target $target$
    \STATE $g_{current} \gets s.target$
    \STATE $succT_N \gets SuccT_M^N(g_{current})$
    \IF {$succT_N \in ||(target) ~\wedge~ type(g_{current}) = ``+" $}
        \STATE $S \gets S \cup \{(g,target)\}$
        \STATE $GR \gets g_{current}$
    \ELSIF {$succT_N \in ||(target) ~\wedge~ type(g_{current}) = ``\times" $}
        \STATE $g_{new} \gets Gateway(``+")$
        \STATE $G \gets G \cup \{g_{new}\}$
        \STATE $S \gets S \setminus \{(source, g_{current}\}$
        \STATE $S \gets S \cup \{(source, g_{new}), (g_{new}, g_{current}), (g_{new}, target)\}$
        \STATE $GR \gets g_{new}$
    \ELSIF {$succ_N \cap ||(target) \gets \emptyset$}
        \STATE $g_{new} \gets Gateway(``\times")$
        \STATE $G \gets G \cup g_{new}$
        \STATE $S \gets S \setminus \{(source, g_{current}\}$
        \STATE $S \gets S \cup \{(source, g_{new}), (g_{new}, g_{current}), (g_{new}, target)\}$
        \STATE $GR \gets g_{new}$
    \ELSE
         \STATE $succ_N \gets Succ_M^N(g_{current})$
        \FORALL{$v \in succ_N$}
            \IF{$v \in G$}
                \STATE $s \gets (g_{current}, v)$
                \STATE $AddNestingSplitGateway(DFG_A, M, \|, \#, GR, s, source, target)$
            \ENDIF
        \ENDFOR
    \ENDIF
\end{algorithmic}
\caption{AddNestingSplitGateway}
\label{algo:split_recursive}
\end{algorithm}

\subsubsection{Join Operation}
\label{subsubsection:join}

Similar to the split operation, the join operation presented in Algorithm~\ref{algo:join} consists of two main parts: adding a simple join gateway and adding a nested join gateway.

\begin{algorithm}[h]  
\footnotesize
\begin{algorithmic}[1]                
\STATE \textbf{Input:} Process Model $M$, gateway reference $GR$, Entity $source$, Entity $target$
\STATE $source_M \gets s.source ~|~ \exists s = (source_M, target) \in S$
\IF {$type(source_M) \notin G$}
    \STATE $g_{new} \gets Gateway(type(GR))$ 
    \STATE $G \gets G \cup \{g_{new}\}$
    \STATE $S \gets S \setminus \{(source, g_{current}\}$
    \STATE $S \gets S \cup \{(source, g_{new}), (source_M, g_{new}), (g_{new}, target)\}$
\ELSE
    \STATE $s \gets (source_M, target)$
    \STATE $AddNestingJoinGateway(M, GR, s, source, target)$
\ENDIF
\end{algorithmic}
\caption{AddJoinGateways}
\label{algo:join}
\end{algorithm}

\begin{figure}[h]
\begin{subfigure}{.4\textwidth}
    \centering
    \includegraphics[width=1\linewidth]{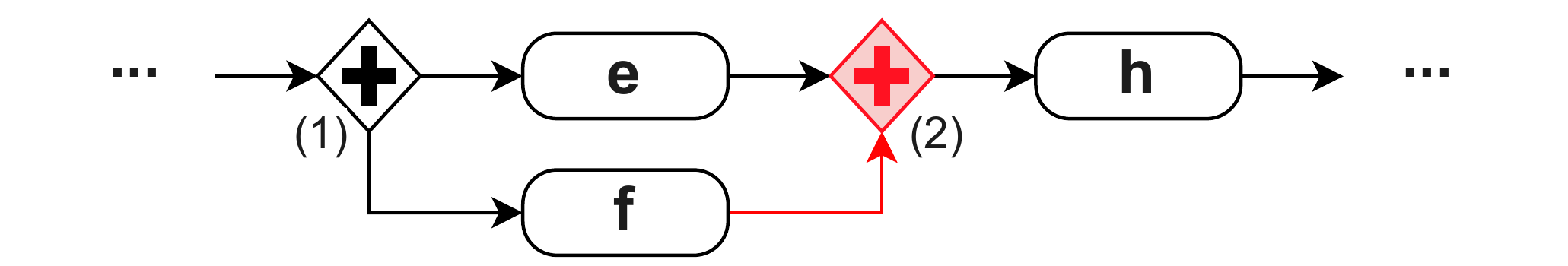}
    \caption{a simple join}
    \label{subfig:join_1}
\end{subfigure}
\begin{subfigure}{.6\textwidth}
    \centering
    \includegraphics[width=1\linewidth]{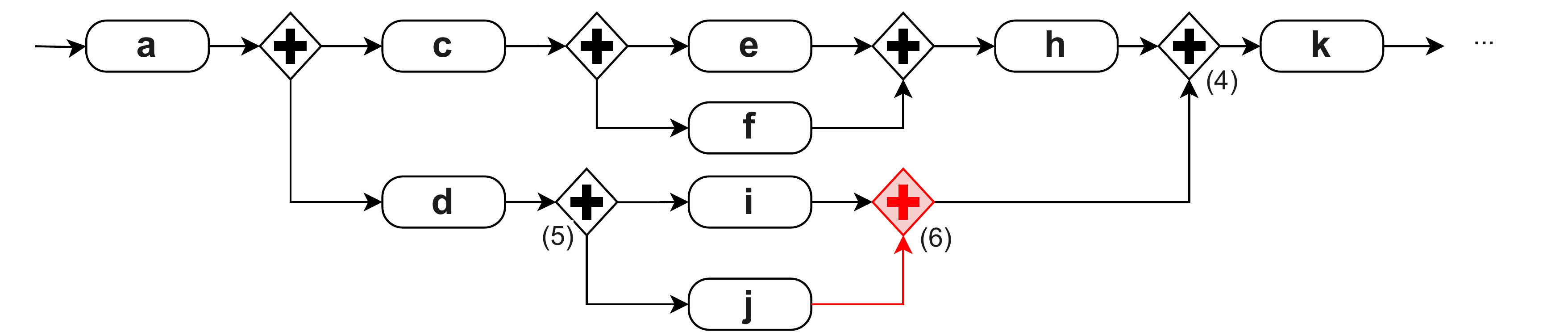}
    \caption{a nested join}
    \label{subfig:join_2}
\end{subfigure}
 \caption{Add join gateways to BPMN.}
 \label{fig:join}
\end{figure}

Figure~\ref{subfig:join_1} illustrates the addition of a simple join. In this example, the dependency relation is $(f, h)$. Since $f$ is a source already added and $h$ is a target already incorporated into the process model, the join algorithm is applied. The first step involves checking the predecessor of $h$, which is $e$ (lines 3-7). As the predecessor is an activity, a join gateway is directly added after $h$. The incoming connections to this gateway are $e$ and $f$, and the outgoing connection is $h$ (lines 9-10). The type of the gateway matches the last gateway added into the BPMN.

The advanced algorithm in Algorithm~\ref{algo:join_recursive} adds a join gateway when a gateway already exists before the target, potentially resulting in nested gateways. It involves three main steps: (i) adding the target as the predecessor of the existing gateway (lines 3-5), (ii) adding a new join gateway after the existing gateway (lines 6-31), and (iii) recalling the complex join algorithm to add a new gateway before the existing one (line 32).

In Figure~\ref{subfig:join_2}, the dependency relation is $(j, k)$, with the last added split being $(5)$. The algorithm checks for a backward path from the selected gateway $(4)$ that intersects the split gateway $(5)$. If all predecessors intersect the split gateway, the temporal relation between the gateways is validated. If they are of the same type, a new relation is added from the target to the selected gateway; otherwise, a new gateway is added after the selected gateway. If at least two predecessors have a relation, a new join gateway matching the split gateway's type is added after these predecessors and before the selected gateway. Otherwise, the nesting join algorithm is applied to all predecessor gateways. In this example, $(4)$ has $h$ and $i$ as predecessors, with a backward path from $i$ to $(5)$. Therefore, a new join gateway of the same type as $(5)$ is added after $i$ (lines 23-30).

\begin{algorithm}[H]  
\footnotesize
\begin{algorithmic}[1]                
\STATE \textbf{Input:} Process Model $M$, gateway reference $GR$, Entity $source$, Entity target $pred$, Entity $target$,  current relation $s$

\STATE $pred_M \gets Pred_M(pred)$
\IF{$\forall v\in pred_M, \exists P_{GR, v}$}
    \IF{$type(pred) == type(GR)$}
        \STATE $S \gets S \cup \{(source,pred)\}$
    \ELSE 
        \STATE $g_{new} \gets Gateway(type(GR))$
        \STATE $G \gets G \cup \{g_{new}\}$
        \STATE $S \gets S \setminus \{s\}$
        \STATE $S \gets S \cup \{(source, g_{new}), (pred, g_{new}), (g_{new},s.target)\}$
    \ENDIF
\ELSIF{$\forall v\in pred_M, \nexists P_{GR, v}$}
    \STATE $g_{new} \gets Gateway(type(GR))$
    \STATE $G \gets G \cup \{g_{new}\}$
    \STATE $S \gets S \setminus \{s\}$
    \STATE $S \gets S \cup \{(source, g_{new}), (pred, g_{new}), (g_{new},s.target)\}$
\ELSIF{$\|\forall v\in pred_M, \exists P_{GR, v}\| \geq 2$}
    \STATE $g_{new} \gets Gateway(type(GR))$
    \STATE $G \gets G \cup \{g_{new}\}$
    \FORALL{$v \in pred_M$}
        \STATE $S \gets S \setminus \{(v,pred)\}$
        \STATE $S \gets S \cup \{(v, g_{new})\}$
    \ENDFOR
    \STATE $S \gets \{(g_{new}, pred)\}$
\ELSE
    \STATE $pred_{new} \gets Pred_M(v) | v \in pred_M$
    \STATE $s = (pred_{new},pred) \in S$
    \IF{$type(pred_{new}) \in G$}
         \STATE $g_{new} \gets Gateway(type(GR))$
        \STATE $G \gets G \cup \{g_{new}\}$
        \STATE $S \gets S \setminus \{s\}$
        \STATE $S \gets S \cup \{(source, g_{new}), (pred_{new}, g_{new}), (g_{new},pred)\}$       
    \ELSE
        \STATE call $AddNestingJoinGateway(M, GR, source, pred_{new}, target, s)$
    \ENDIF

\ENDIF
\STATE $GR \gets null$
\end{algorithmic}
\caption{AddNestingJoinGateway}
\label{algo:join_recursive}
\end{algorithm}

\subsection{Loop Construction}
\label{subsec:loop_construction}

\begin{figure}[h]
\includegraphics[width=0.6\linewidth]{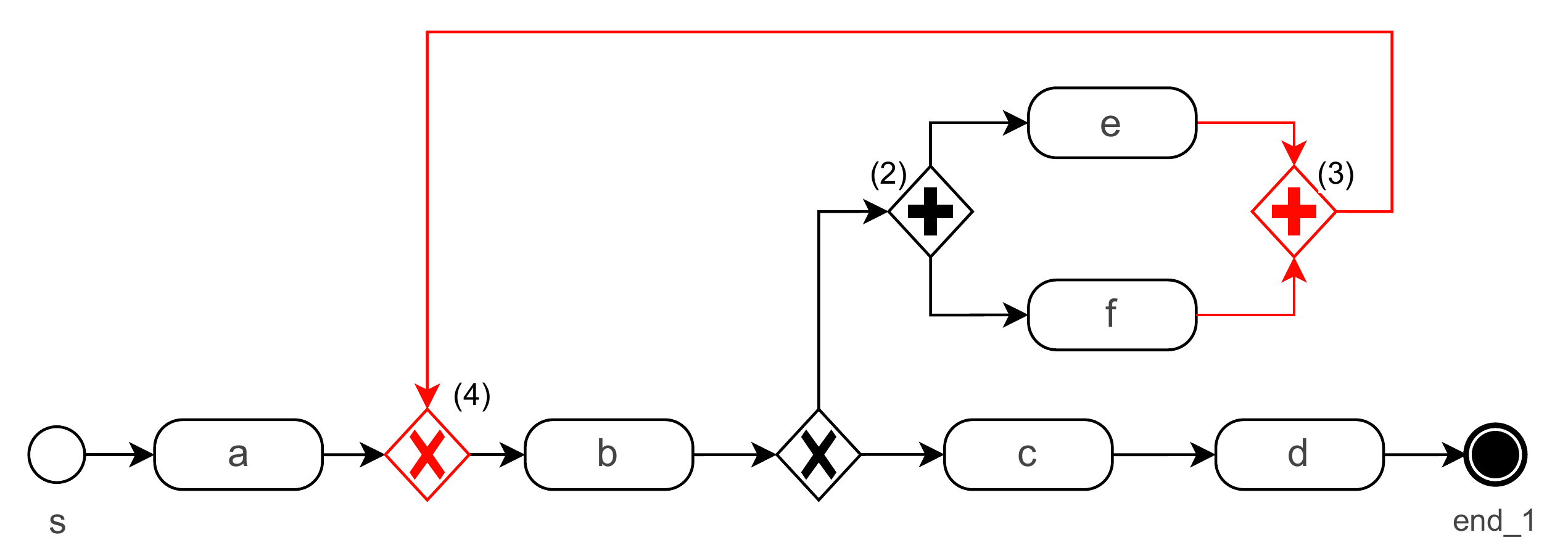}
\centering
\caption{Complete a loop block and add it to the BPMN model.}
\label{fig:loop_construction}
\end{figure}

We start in the loop construction by the block construction. During the generation of the BPMN, it is not possible to establish all relationships because the loop relation is removed, resulting in some incomplete blocks. For example, after removing the entities in red in Figure~\ref{fig:loop_construction}, the result is contracted before adding the loop. In the loop block construction algorithm (Algorithm~\ref{algo:block_construction}), the goal is to complete the loop block (gateway (3)) before adding the loops as described in algorithm~\ref{algo:loop}.

\begin{algorithm}[h]
\footnotesize
\begin{algorithmic}[1]
\STATE \textbf{Input:} Acyclic DFG $DFG_A$, Process Model $M=(N,G,S)$, Merged loops $ML$, parallel vertices $\|$, exclusive vertices $\#$

\FORALL{$l = (sources,targets) \in ML$}
    \IF{$\|sources\| \neq 1 ~\wedge~ \nexists (u,v) \in S ~|~ (v\in N ~\wedge~ u\in sources)$}
            \STATE $g_{split} \gets LCA_{M}(sources)$
            \STATE $G \gets G \cup \{g_{join}\} ~|~ type(g_{join}) = type(g_{split})$
            \STATE $S \gets S \cup \{(s, g_{join}\} ~|~ s \in sources$
            
    \ENDIF

         \IF{$\|targets\| \neq 1 ~\wedge~ \nexists (v,u) \in S ~|~ (v\in N ~\wedge~ u\in targets)$}
            \STATE $ N \gets N \cup \{temp\}$
            \FORALL{$t \in targets$}
                \STATE call $AddSplitGateways(DFG_A, M, \|, \#, temp, t, \emptyset)$ 
            \ENDFOR
            \STATE $ N \gets N \setminus \{temp\}$
        \ENDIF    
\ENDFOR

\end{algorithmic}
\caption{Loop Block Construction}
\label{algo:block_construction}
\end{algorithm}

\begin{figure}[h]
\includegraphics[width=0.75\linewidth]{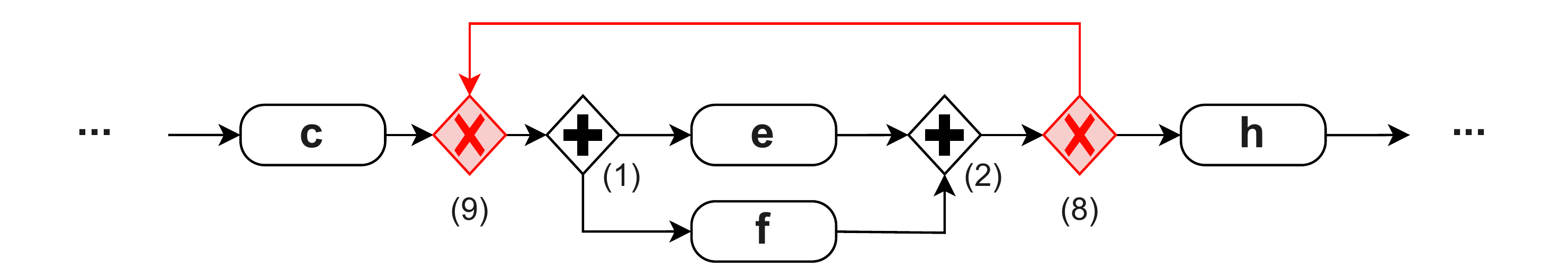}
\centering
\caption{Add a loop to the process model.}
\label{fig:loop}
\end{figure}

Since in the example used in the paper, the blocks of {e, f} are completed where the source block is gateway (1) and the target block is gateway (2) (cf. Figure ~\ref{fig:loop}), we add other example (cf. Figure~\ref{fig:loop_construction}) to well explain the algorithm, where only one loop exists, which is $(\{e,f\},\{b\})$. Since there are two sources, $\{e,f\}$, it is necessary to validate the completion of their block (line 3). If the block is completed, no action is required; otherwise, the block should be constructed based on the split. To accomplish this, by applying the lowest common ancestor algorithm~\cite{DBLP:journals/jal/BenderFPSS05} on the process model, the split gateway can be identified as the last ancestor $e$ and $f$ (line 4). Subsequently, a new join gateway should be added with the same type of split(lines 5-6).

Next, the targets are checked to determine if they construct all the blocks. If more than one target exists, the existence of gateways before the targets is validated (line 7). If gateways do not exist, a temporary activity is added to construct this block by adding connections and applying the split operation from the temporary activity to the targets (line 8-10). Finally, this activity is removed (line 11), and the process continues to the final algorithms to add loops to the process model~\ref{algo:loop}.

The final step involves adding the loop between the source and target blocks presented in Algorithm~\ref{algo:loop}. A block is considered an activity if there is one element for the source or target, and it is considered a gateway if there is more than one source or target. The process iterates through all the Merged Loop results, leading to four scenarios: (i) If both source and target blocks have a sequence flow to other entities, a new gateway is added after the source block and another before the target block. These two gateways are then connected (line 7). (ii) If the source block has a sequence flow but the target block does not, a new gateway is added after the source block, followed by a connection from the new gateway to the target block (line 9). (iii) If the target block has a sequence flow but the source block does not, a new gateway is added before the target block, followed by a connection from the source block to the gateway (line 12). (iv) If neither the source nor the target block has a sequence flow, a simple connection is created between these blocks (line 14).

\begin{algorithm}[h]
\footnotesize
\begin{algorithmic}[1]
\STATE \textbf{Input:} Process Model $M$, merge loops $ML$

\FORALL{$l \in ML$}
    \STATE $sourceBlock \gets \bigcap \{Succ_M(s) | s \in l_s\}$ if $|l_s| > 1$ else $l_s$
    \STATE $targetBlock \gets LCA_M(l_t)$ if $|l_t| > 1$ else $l_t$
    \IF{$\exists (sourceBlock,s) \in S$}
        \IF{$\exists (s,targetBlock) \in S$}
            \STATE $AddTwoLoopGateways(sourceBlock,targetBlock)$
        \ELSE
            \STATE $AddLoopAfterSourceBlock(sourceBlock,targetBlock)$
        \ENDIF
    \ELSE
        \IF{$\exists (s,targetBlock) \in S$}
            \STATE $AddLoopBeforeTargetBlock(sourceBlock,targetBlock)$
        \ELSE
            \STATE $S \gets S \cup \{(sourceBlock,targetBlock)\}$
        \ENDIF
    \ENDIF
\ENDFOR

\end{algorithmic}
\caption{Loop Operation}
\label{algo:loop}
\end{algorithm}

In the paper example, Figure~\ref{fig:loop} represents the loop operation where the $\{e,f\}$ block is in a loop, meaning this block is both the source and the target, with a gateway as both the source and target blocks. The source block is gateway (2), and the target block is gateway (1). These blocks are connected; therefore, new gateways are added, one after the source (8) and another before the target (9), and a connection from (8) to (9) is added (line 14 in the algorithm).

In other case (Figure~\ref{fig:loop_construction}), after constructing the loop block (adding the gateway $(3)$), the source block is disconnected, but $b$ which is the target block is connected to other entities. Therefore, a new $\times$ gateway $(4)$ is added before $b$, with a connection from $(3)$ to $(4)$ (lines 11-12 in the algorithm).



\section{Evaluation}
\label{sec:evaluation}
We implemented Bonita Miner (hereafter BM) as a standalone Java application~\footnote{Available at: \url{https://github.com/NourEldin-Ali/BonitaMiner}}. The tool takes as input an event log in XES format and produces a BPMN process model as output. Using this implementation, we conducted an empirical comparison of BM against three existing methods using a set of publicly available logs.

\subsection{Datasets}
We used a collection of real-life event logs, including logs from the annual Business Process Intelligence Challenge (BPIC), as well as other logs such as the Road Traffic Fines Management Process (RTFMP) and the SEPSIS Cases log
as presented in Table~\ref{table:log}.
For the BPIC logs, we applied a filtering method to remove infrequent traces (5\%) before applying each of the discovery methods. The dataset is heterogeneous in the number of traces (44 to 150,370), in the number of distinct traces (15 to 2023), in the number of event classes (11 to 55) and in the trace length (2 to 185 events).


\begin{longtable}{|c|c|c|c|c|ccc|}
\hline
\multirow{2}{*}{Model} & \multirow{2}{*}{Total traces} & \multirow{2}{*}{Distinct traces} & \multirow{2}{*}{Total events} & \multirow{2}{*}{Distinct events} & \multicolumn{3}{c|}{Trace length}                         \\ \cline{6-8} 
                       &                               &                                  &                               &                                  & \multicolumn{1}{c|}{Min} & \multicolumn{1}{c|}{Avr} & Max \\ \hline
\endfirsthead
\multicolumn{8}{c}%
{{\bfseries Table \thetable\ continued from previous page}} \\
\endhead
$BPIC 2012_f$            & 9333                          & 612                              & 95348                         & 23                               & \multicolumn{1}{c|}{3}   & \multicolumn{1}{c|}{10}  & 55  \\ \hline
$BPIC 2013_{cp_f}$      & 1236                          & 35                               & 4343                          & 4                                & \multicolumn{1}{c|}{2}   & \multicolumn{1}{c|}{3}   & 13  \\ \hline
$BPIC 2013_{inc_f}$      & 5620                          & 178                              & 31740                         & 4                                & \multicolumn{1}{c|}{2}   & \multicolumn{1}{c|}{5}   & 18  \\ \hline
$BPIC 2013_{op_f}$       & 693                           & 30                               & 1569                          & 3                                & \multicolumn{1}{c|}{1}   & \multicolumn{1}{c|}{2}   & 9   \\ \hline
$BPIC 2015_{1_f}$          & 44                            & 15                               & 473                           & 55                               & \multicolumn{1}{c|}{2}   & \multicolumn{1}{c|}{10}  & 38  \\ \hline
$BPIC 2015_{2_f}$          & 8                             & 4                                & 88                            & 15                               & \multicolumn{1}{c|}{7}   & \multicolumn{1}{c|}{11}  & 13  \\ \hline
$BPIC 2015_{3_f}$          & 76                            & 16                               & 650                           & 51                               & \multicolumn{1}{c|}{3}   & \multicolumn{1}{c|}{8}   & 38  \\ \hline
$BPIC 2015_{4_f}$          & 7                             & 3                                & 73                            & 14                               & \multicolumn{1}{c|}{4}   & \multicolumn{1}{c|}{10}  & 13  \\ \hline
$BPIC 2015_{5_f}$          & 6                             & 3                                & 64                            & 18                               & \multicolumn{1}{c|}{5}   & \multicolumn{1}{c|}{10}  & 15  \\ \hline
$BPIC 2017_f$            & 17602                         & 2023                             & 501041                        & 24                               & \multicolumn{1}{c|}{10}  & \multicolumn{1}{c|}{28}  & 63  \\ \hline
$BPIC 2017_{o_f}$        & 40754                         & 6                                & 184037                        & 7                                & \multicolumn{1}{c|}{3}   & \multicolumn{1}{c|}{4}   & 5   \\ \hline
$BPIC 2020_{DD_f}$       & 9899                          & 8                                & 51479                         & 10                               & \multicolumn{1}{c|}{1}   & \multicolumn{1}{c|}{5}   & 9   \\ \hline
$BPIC 2020_{ID_f}$       & 5989                          & 293                              & 65739                         & 32                               & \multicolumn{1}{c|}{3}   & \multicolumn{1}{c|}{10}  & 24  \\ \hline
$BPIC 2020_{PL_f}$       & 5972                          & 385                              & 65575                         & 44                               & \multicolumn{1}{c|}{3}   & \multicolumn{1}{c|}{10}  & 30  \\ \hline
$BPIC 2020_{PTC_f}$      & 1991                          & 94                               & 17040                         & 29                               & \multicolumn{1}{c|}{1}   & \multicolumn{1}{c|}{8}   & 16  \\ \hline
$BPIC 2020_{RP_f}$       & 6521                          & 11                               & 34328                         & 11                               & \multicolumn{1}{c|}{1}   & \multicolumn{1}{c|}{5}   & 11  \\ \hline
$RTFMP$                  & 150370                        & 231                              & 561470                        & 11                               & \multicolumn{1}{c|}{2}   & \multicolumn{1}{c|}{4}   & 20  \\ \hline
$SEPSIS$                 & 1050                          & 846                              & 15214                         & 16                               & \multicolumn{1}{c|}{3}   & \multicolumn{1}{c|}{14}  & 185 \\ \hline
\caption{Statistics of the event logs employed}
\label{table:log}\\
\end{longtable}

\begin{figure}[h]
    \centering
    \begin{subfigure}{0.48\linewidth}
        \centering
        \includegraphics[width=\linewidth]{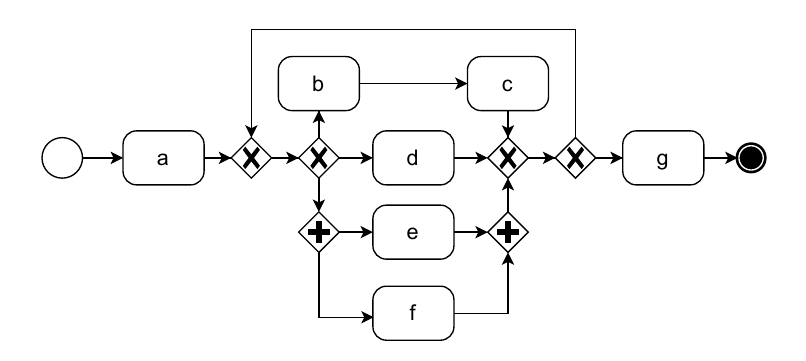}
        \caption{(L1): BPMN with loop block, the loop is for a block that contains nesting gateways and parallelism.}
        \label{fig:sub:l1}
    \end{subfigure}
    \hfill
    \begin{subfigure}{0.48\linewidth}
        \centering
        \includegraphics[width=\linewidth]{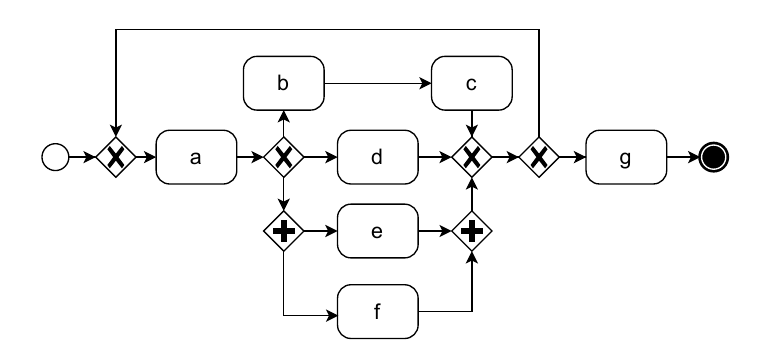}
        \caption{(L2): BPMN with loop block, the loop is from different sources to one target that contains nesting gateways and parallelism.}
        \label{fig:sub:l2}
    \end{subfigure}

     \begin{subfigure}{0.48\linewidth}
        \centering
        \includegraphics[width=\linewidth]{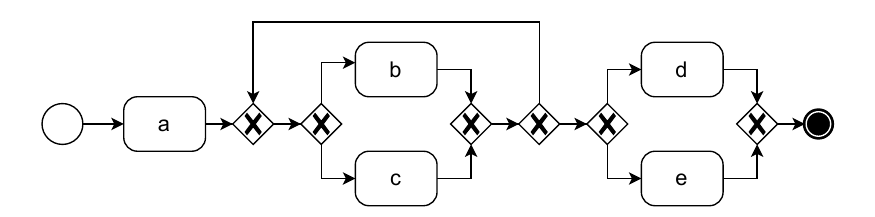}
        \caption{(L3): Loops block on XOR gateway followed by split gateway.}
        \label{fig:sub:l3}
    \end{subfigure}
    \hfill
    \begin{subfigure}{0.48\linewidth}
        \centering
        \includegraphics[width=\linewidth]{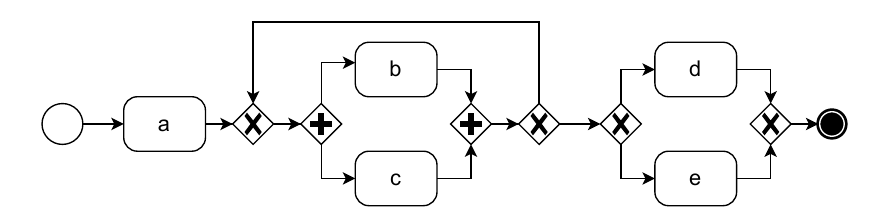}
        \caption{(L4): Loops block on AND gateway followed by split gateway.}
        \label{fig:sub:l4}
    \end{subfigure}
    \caption{Different Loops Patterns}
    \label{fig:bpmn_log}
\end{figure}

In addition to these logs, we generated event logs for the BPMN models containing loops, as shown in Figure~\ref{fig:bpmn_log}, which are the focus of this paper. These patterns are used for different loop block, like multi source to multi target in BPMN (L1)~\ref{fig:sub:l1}, or multi source one target in BPMN (L2)~\ref{fig:sub:l2}, also the representation of the Block loops followed by split like in L3~\ref{fig:sub:l3} and L4 ~\ref{fig:sub:l4}.

\subsection{Experimental Setup}

We selected three state-of-the-art discovery methods as baselines: Inductive Miner (IM), Heuristics Miner (HM), and Split Miner (SM). IM and HM were implemented using the PM4PY library~\cite{pm4py}.

For the evaluation of default parameters, we assessed the quality of the produced models across various quality dimensions. These included fitness, precision, and F-score as proxies for accuracy, as well as generalization, size, control-flow complexity (CFC), and structuredness (struct.) as proxies for complexity~\cite{ProcessMiningBook}, along with execution time in seconds, excluding the time required to load the event log into memory. The size metric represents the number of elements in the process model, such as nodes (tasks, events, gateways) and edges, while CFC measures the structural complexity of the control-flow, considering constructs like splits and joins (XOR, AND). Each metric was computed on BPMN models, except for fitness, precision, and generalization, which were measured on Petri nets due to the limitations of the measuring tools, which work exclusively on Petri nets. We converted BPMN models to Petri nets using the PM4PY library and also used PM4PY to calculate fitness, precision, and generalization.

All experiments were conducted on a system equipped with an 11th Gen Intel Core i7 processor running at 2.30 GHz, paired with 16 GB of RAM.

\subsection{Results \& Discussion}
The evaluation of BM revealed its superior performance across multiple dimensions. The results, summarized in Table~\ref{table:results}, show that BM consistently outperformed baseline methods across synthetic and real-life datasets. For instance, BM achieved perfect fitness (1.00) on most synthetic logs (L1–L4), indicating its robustness in reproducing observed behaviors. In real-life logs such as BPIC 2017, BM maintained competitive fitness (0.93) and F-score (0.94), striking a commendable balance between accuracy and simplicity.

\begin{longtable}{cccccccccc}
\textbf{Log name}                  & Discovery Method & \multicolumn{3}{c}{Distinct traces}           & Generalization & \multicolumn{3}{c}{Complexity}            & Time(s)        \\ \cline{3-5} \cline{7-9}
\endfirsthead
\multicolumn{10}{c}%
{{\bfseries Table \thetable\ continued from previous page}} \\
\endhead
                                   &                  & Fitness       & Precision     & F-score       &                & Size        & CFC         & Struct.       &                \\ \hline
\multirow{4}{*}{L1}                & BM               & \textbf{1.00} & 0.91          & \textbf{0.95} & \textbf{0.97}  & \textbf{15} & \textbf{6}  & \textbf{1.00} & 0.091          \\
                                   & SM               & 0.89          & 0.87          & 0.88          & 0.95           & 23          & 13          & 0.65          & \textbf{0.01}  \\
                                   & IM               & 0.78          & \textbf{1.00} & 0.88          & 0.88           & 39          & 30          & 0.103         & 0.19           \\
                                   & HM               & \textbf{1.00} & 0.6           & 0.75          & 0.95           & 31          & 21          & 1.00          & 0.21           \\ \hline
\multirow{4}{*}{L2}                & BM               & \textbf{1.00} & 0.96          & \textbf{0.98} & \textbf{0.97}  & \textbf{15} & \textbf{6}  & \textbf{1.0}  & \textbf{0.05}  \\
                                   & SM               & 0.81          & \textbf{1.00} & 0.90          & 0.95           & \textbf{15} & 7           & 0.80          & 0.09           \\
                                   & IM               & 0.96          & 0.96          & 0.96          & 0.89           & 17          & 12          & 0.176         & 0.24           \\
                                   & HM               & \textbf{1.00} & 0.96          & \textbf{0.98} & 0.96           & \textbf{15} & \textbf{6}  & \textbf{1.0}  & 0.24           \\ \hline
\multirow{4}{*}{L3}                & BM               & \textbf{1.00} & 0.99          & \textbf{0.99} & \textbf{0.96}  & \textbf{12} & \textbf{5}  & \textbf{1.0}  & \textbf{0.053} \\
                                   & SM               & 0.82          & 0.98          & 0.90          & 0.95           & 15          & 7           & 1.0           & 0.06           \\
                                   & IM               & 0.79          & \textbf{1.00} & 0.88          & 0.94           & 21          & 11          & 0.14          & 0.16           \\
                                   & HM               & \textbf{1.00} & 0.84          & 0.91          & 0.95           & 19          & 11          & 1.00          & 0.15           \\ \hline
\multirow{4}{*}{L4}                & BM               & \textbf{1.00} & 0.99          & \textbf{0.99} & \textbf{0.97}  & \textbf{12} & \textbf{4}  & \textbf{1.0}  & 0.09           \\
                                   & SM               & \textbf{1.00} & 0.84          & 0.91          & 0.96           & 15          & 7           & \textbf{1.0}  & \textbf{0.06}  \\
                                   & IM               & 0.79          & \textbf{1.00} & 0.88          & 0.95           & 21          & 10          & 0.14          & 0.17           \\
                                   & HM               & \textbf{1.00} & 0.84          & 0.91          & 0.96           & 15          & 7           & \textbf{1.00} & 0.16           \\ \hline
\multirow{4}{*}{$BPIC 2012_f$}       & BM               & 0.93          & \textbf{0.82} & \textbf{0.87} & 0.85           & \textbf{48} & \textbf{34} & 0.67          & 4.2            \\
                                   & SM               & 0.82          & \textbf{0.82} & 0.82          & 0.86           & 95          & 62          & 0.32          & \textbf{1.46}  \\
                                   & IM               & \textbf{1.00} & 0.28          & 0.44          & \textbf{0.96}  & 86          & 58          & \textbf{1.00} & 3.69           \\
                                   & HM               & 0.98          & 0.36          & 0.52          & 0.92           & 49          & 45          & -             & 3.29           \\ \hline
\multirow{4}{*}{$BPIC 2013_{cp_f}$}  & BM               & 0.91          & 0.88          & 0.89          & 0.89           & \textbf{12} & \textbf{6}  & \textbf{1.00} & 0.18           \\
                                   & SM               & 0.85          & \textbf{0.94} & 0.90          & 0.88           & \textbf{12} & \textbf{6}  & \textbf{1.00} & 0.04           \\
                                   & IM               & \textbf{1.00} & 0.63          & 0.77          & \textbf{0.91}  & 18          & 11          & 1.00          & 0.27           \\
                                   & HM               & \textbf{1.00} & 0.90          & \textbf{0.94} & 0.89           & 11          & 6           & \textbf{0.73} & \textbf{0.28}  \\ \hline
\multirow{4}{*}{$BPIC 2013_{inc_f}$} & BM               & 0.87          & 0.98          & 0.92          & 0.90           & \textbf{13} & 11          & -             & 0.59           \\
                                   & SM               & 0.98          & \textbf{0.99} & \textbf{0.98} & \textbf{0.95}  & 16          & \textbf{10} & \textbf{1.00} & \textbf{0.08}  \\
                                   & IM               & \textbf{1.00} & 0.61          & 0.76          & 0.94           & 16          & \textbf{10} & \textbf{1.00} & 1.9            \\
                                   & HM               & \textbf{1.00} & 0.96          & 0.98          & 0.94           & \textbf{13} & \textbf{10} & 0.85          & 2.05           \\ \hline
\multirow{4}{*}{$BPIC 2013_{op_f}$}  & BM               & 0.98          & \textbf{0.99} & \textbf{0.99} & \textbf{0.93}  & \textbf{11} & \textbf{7}  & 0.64          & 0.03           \\
                                   & SM               & 0.83          & 0.96          & 0.89          & \textbf{0.93}  & 13          & \textbf{7}  & \textbf{1.00} & \textbf{0.02}  \\
                                   & IM               & \textbf{1.00} & 0.75          & 0.86          & 0.91           & 23          & 16          & \textbf{1.00} & 0.13           \\
                                   & HM               & \textbf{1.00} & 0.97          & \textbf{0.99} & 0.90           & 11          & 9           & 0.91          & 0.11           \\ \hline
\multirow{4}{*}{$BPIC 2015_{1_f}$}     & BM               & 0.82          & \textbf{0.91} & \textbf{0.87} & 0.5            & \textbf{90} & \textbf{34} & 0.36          & \textbf{0.03}  \\
                                   & SM               & 0.77          & 0.85          & 0.81          & 0.49           & 172         & 99          & 0.23          & \textbf{0.05}  \\
                                   & IM               & 0.92          & 0.79          & 0.85          & 0.48           & 116         & 63          & 0.31          & 0.16           \\
                                   & HM               & \textbf{1.00} & 0.30          & 0.46          & \textbf{0.56}  & 121         & 58          & \textbf{1.00} & 0.25           \\ \hline
\multirow{4}{*}{$BPIC 2015_{2_f}$}     & BM               & 0.92          & \textbf{0.92} & 0.92          & 0.52           & \textbf{28} & 12          & -             & \textbf{0.01}  \\
                                   & SM               & 0.97          & 0.88          & 0.92          & 0.46           & 33          & 17          & 0.15          & 0.04           \\
                                   & IM               & \textbf{1.00} & 0.64          & 0.78          & \textbf{0.54}  & 30          & \textbf{11} & \textbf{1.00} & 0.1            \\
                                   & HM               & \textbf{1.00} & 0.88          & \textbf{0.94} & 0.49           & 33          & 16          & 0.15          & 0.1            \\ \hline
\multirow{4}{*}{$BPIC 2015_{3_f}$}     & BM               & 0.84          & \textbf{0.92} & \textbf{0.88} & 0.53           & \textbf{82} & \textbf{30} & 0.28          & \textbf{0.05}  \\
                                   & SM               & 0.72          & 0.68          & 0.70          & 0.49           & 117         & 52          & 0.20          & 0.15           \\
                                   & IM               & \textbf{1.00} & 0.49          & 0.66          & \textbf{0.56}  & 102         & 46          & \textbf{1.00} & 0.18           \\
                                   & HM               & 0.90          & 0.86          & \textbf{0.88} & 0.52           & 91          & 39          & 0.25          & 0.16           \\ \hline
\multirow{4}{*}{$BPIC 2015_{4_f}$}     & BM               & 0.89          & \textbf{0.92} & \textbf{0.90} & \textbf{0.55}  & \textbf{22} & \textbf{6}  & \textbf{1.00} & \textbf{0.01}  \\
                                   & SM               & 0.89          & \textbf{0.92} & \textbf{0.90} & \textbf{0.55}  & \textbf{22} & \textbf{6}  & \textbf{1.00} & 0.02           \\
                                   & IM               & \textbf{1.00} & 0.71          & 0.83          & 0.53           & 26          & 8           & \textbf{1.00} & 0.1            \\
                                   & HM               & 0.88          & \textbf{0.92} & \textbf{0.90} & \textbf{0.55}  & \textbf{22} & \textbf{6}  & \textbf{1.00} & 0.1            \\ \hline
\multirow{4}{*}{$BPIC 2015_{5_f}$}     & BM               & 0.96          & \textbf{0.89} & \textbf{0.92} & 0.38           & \textbf{33} & \textbf{15} & 0.67          & \textbf{0.01}  \\
                                   & SM               & \textbf{1.00} & 0.85          & \textbf{0.92} & \textbf{0.41}  & 36          & 16          & 0.64          & 0.04           \\
                                   & IM               & \textbf{1.00} & 0.47          & 0.64          & \textbf{0.41}  & 36          & 16          & \textbf{1.00} & 0.12           \\
                                   & HM               & \textbf{1.00} & 0.85          & \textbf{0.92} & 0.38           & 36          & 18          & 0.61          & 0.12           \\ \hline
\multirow{4}{*}{$BPIC 2017_f$}       & BM               & 0.93          & 0.96          & \textbf{0.94} & 0.80           & \textbf{54} & 50          & 0.69          & 15.68          \\
                                   & SM               & 0.83          & \textbf{1.00} & 0.91          & 0.80           & 142         & 116         & 0.23          & \textbf{0.21}  \\
                                   & IM               & 0.94          & 0.85          & 0.89          & 0.92           & 62          & \textbf{47} & 0.45          & 25.24          \\
                                   & HM               & \textbf{1.00} & 0.22          & 0.36          & \textbf{0.94}  & 91          & 69          & \textbf{1.00} & 30.92          \\ \hline
\multirow{4}{*}{$BPIC 2017_{o_f}$}   & BM               & \textbf{1.00} & \textbf{1.00} & \textbf{1.00} & \textbf{0.99}  & 15          & 8           & 0.27          & 5.2            \\
                                   & SM               & \textbf{1.00} & \textbf{1.00} & \textbf{1.00} & \textbf{0.99}  & 15          & 8           & 0.27          & \textbf{0.11}  \\
                                   & IM               & \textbf{1.00} & 0.80          & 0.89          & 0.99           & \textbf{14} & \textbf{7}  & \textbf{1.00} & 11.02          \\
                                   & HM               & 0.91          & \textbf{1.00} & 0.95          & 0.79           & 18          & 10          & 0.22          & 11.79          \\ \hline
\multirow{4}{*}{$BPIC 2020_{DD_f}$}  & BM               & \textbf{1.00} & \textbf{0.99} & \textbf{1.00} & \textbf{0.97}  & 19          & 10          & 0.58          & 2.5            \\
                                   & SM               & \textbf{1.00} & \textbf{0.99} & \textbf{1.00} & \textbf{0.97}  & 21          & 10          & 0.57          & \textbf{0.05}  \\
                                   & IM               & \textbf{1.00} & 0.70          & 0.82          & \textbf{0.97}  & 23          & 13          & \textbf{1.00} & 1.9            \\
                                   & HM               & 0.91          & 0.99          & 0.95          & 0.90           & \textbf{18} & \textbf{8}  & 0.28          & 2.09           \\ \hline
\multirow{4}{*}{$BPIC 2020_{ID_f}$}  & BM               & 0.89          & 0.93          & 0.91          & 0.84           & \textbf{65} & \textbf{47} & 0.19          & 2.96           \\
                                   & SM               & 0.85          & 0.75          & 0.79          & 0.78           & 149         & 125         & 0.05          & \textbf{0.37}  \\
                                   & IM               & \textbf{1.00} & 0.25          & 0.40          & \textbf{0.93}  & 112         & 76          & \textbf{1.00} & 3.26           \\
                                   & HM               & 0.91          & 0.94          & \textbf{0.93} & 0.72           & 100         & 89          & 0.10          & 3.44           \\ \hline
\multirow{4}{*}{$BPIC 2020_{PL_f}$}  & BM               & 0.84          & 0.96          & 0.89          & 0.81           & \textbf{85} & \textbf{60} & \textbf{0.51} & 2.99           \\
                                   & SM               & 0.75          & 0.91          & 0.82          & 0.79           & 229         & 195         & 0.12          & \textbf{0.55}  \\
                                   & IM               & \textbf{1.00} & 0.09          & 0.17          & \textbf{0.90}  & -           & -           & -             & 3.95           \\
                                   & HM               & 0.89          & \textbf{0.97} & \textbf{0.93} & 0.74           & 118         & 103         & 0.16          & 3.39           \\ \hline
\multirow{4}{*}{$BPIC 2020_{PTC_f}$} & BM               & 0.93          & 0.91          & \textbf{0.92} & 0.77           & \textbf{62} & \textbf{44} & 0.19          & 0.96           \\
                                   & SM               & 0.78          & 0.76          & 0.77          & 0.72           & 163         & 128         & 0.01          & \textbf{0.28}  \\
                                   & IM               & \textbf{0.98} & 0.11          & 0.20          & \textbf{0.89}  & 100         & 73          & \textbf{1.00} & 0.95           \\
                                   & HM               & 0.91          & \textbf{0.94} & \textbf{0.92} & 0.68           & 88          & 76          & 0.11          & 0.9            \\ \hline
\multirow{4}{*}{$BPIC 2020_{RP_f}$}  & BM               & \textbf{1.00} & \textbf{0.99} & \textbf{0.99} & 0.96           & 21          & 11          & 0.48          & 1.96           \\
                                   & SM               & \textbf{1.00} & \textbf{0.99} & \textbf{0.99} & \textbf{0.97}  & 24          & 11          & 0.46          & 0.05           \\
                                   & IM               & \textbf{1.00} & 0.78          & 0.88          & \textbf{0.97}  & 24          & 15          & \textbf{1.00} & 1.5            \\
                                   & HM               & 0.92          & \textbf{0.99} & 0.95          & 0.90           & \textbf{20} & \textbf{9}  & 0.10          & 1.62           \\ \hline
\multirow{4}{*}{$RTFMP_f$}           & BM               & 0.98          & \textbf{1.00} & \textbf{0.98} & 0.99           & \textbf{13} & \textbf{6}  & 0.46          & 13.26          \\
                                   & SM               & 0.88          & 0.99          & 0.93          & \textbf{1.00}  & 21          & 10          & 0.29          & \textbf{0.29}  \\
                                   & IM               & 0.92          & \textbf{1.00} & 0.96          & 0.84           & 20          & 14          & 0.15          & 24.03          \\
                                   & HM               & \textbf{1.00} & 0.66          & 0.80          & \textbf{1.00}  & 22          & 15          & \textbf{1.00} & 21.64          \\ \hline
\multirow{4}{*}{$SEPSIS_f$}          & BM               & 0.79          & \textbf{0.99} & \textbf{0.88} & 0.76           & \textbf{26} & \textbf{16} & 0.41          & 0.35           \\
                                   & SM               & 0.92          & 0.70          & 0.79          & 0.74           & 34          & 21          & 0.71          & \textbf{0.02}  \\
                                   & IM               & 0.90          & 0.83          & 0.86          & 0.72           & 47          & 39          & 0.23          & 0.13           \\
                                   & HM               & \textbf{1.00} & 0.60          & 0.75          & \textbf{0.86}  & 40          & 25          & \textbf{1.00} & 0.12          
\label{table:results}\\
\caption{Evaluation results}
\end{longtable}

The generalization capabilities of BM stand out as a key strength. Across the majority of datasets, BM consistently achieved scores that matched or surpassed those of the baseline methods, demonstrating its ability to generalize beyond observed behaviors without succumbing to overgeneralization. For instance, in the RTFMP log, BM achieved an impressive generalization score of 0.99, while maintaining minimal structural complexity, highlighting its superior adaptability and precision.

Regarding model simplicity, BM frequently outperformed its competitors by generating smaller, more interpretable models with lower structural complexity. A notable example is the SEPSIS log, where BM produced a model with only 26 elements and a CFC of 16, significantly enhancing interpretability compared to SM, IM, and HM. Similarly, in synthetic loops such as L3 and L4, BM maintained high fitness and precision while preserving structural clarity, further showcasing its effectiveness in handling complex loop structures.

In terms of execution time, SM remains the fastest method for real-life event logs. However, BM demonstrates improved efficiency over IM and HM, striking a commendable balance between performance and computational demand.

Overall, the empirical results underscore BM’s ability to deliver accurate, interpretable, and generalizable models. While other methods, such as SM and HM, exhibit specific strengths in certain scenarios, BM provides a comprehensive solution, excelling particularly in handling loops and generating well-rounded models across diverse datasets.

\section{Conclusion \& Future Work}
\label{sec:conclusion}

This paper introduces Bonita Miner, a novel approach to process discovery that effectively balances key metrics—fitness, precision, generalization, and simplicity—even when dealing with complex loop structures. By leveraging the Depth-First Algorithm (DFA) for split and join gateway construction and advanced filtering of Directly-Follows Graphs (DFG), Bonita Miner simplifies the process model generation while maintaining behavioral accuracy. The proposed method addresses the limitations of existing techniques by producing simpler, more interpretable models that are robust to the intricacies of real-world processes, such as concurrency and loops. Empirical results demonstrate its superiority over state-of-the-art methods, particularly in its ability to generate behaviorally accurate and structurally simpler models.

Future enhancements to Bonita Miner could focus on several areas of development. One area is the optimization of execution time, as reducing the computational overhead for processing large-scale event logs would make the method more practical for industry-scale applications. Another promising direction is the incorporation of automatic deadlock detection and prevention mechanisms. This would not only enhance the robustness of the process models but also ensure their operational soundness in real-world scenarios.

\bibliographystyle{splncs04}
\bibliography{bibliography}

\begin{thebibliography}{10}
\providecommand{\url}[1]{\texttt{#1}}
\providecommand{\urlprefix}{URL }
\providecommand{\doi}[1]{https://doi.org/#1}

\bibitem{ProcessMiningBook}
van~der Aalst, W.M.P.: Process Mining - Data Science in Action, Second Edition. Springer (2016)

\bibitem{DBLP:journals/tkde/AalstWM04}
van~der Aalst, W.M.P., Weijters, T., Maruster, L.: Workflow mining: Discovering process models from event logs. {IEEE} Trans. Knowl. Data Eng.  \textbf{16}(9),  1128--1142 (2004)

\bibitem{DBLP:conf/er/AugustoCDRB16}
Augusto, A., Conforti, R., Dumas, M., Rosa, M.L., Bruno, G.: Automated discovery of structured process models: Discover structured vs. discover and structure. In: Conceptual Modeling - 35th International Conference, {ER} 2016, Gifu, Japan, November 14-17, 2016, Proceedings. Lecture Notes in Computer Science, vol.~9974, pp. 313--329 (2016)

\bibitem{process-discovery-review}
Augusto, A., Conforti, R., Dumas, M., Rosa, M.L., Maggi, F.M., Marrella, A., Mecella, M., Soo, A.: Automated discovery of process models from event logs: Review and benchmark. {IEEE} Trans. Knowl. Data Eng.  \textbf{31}(4),  686--705 (2019)

\bibitem{splitminer}
Augusto, A., Conforti, R., Dumas, M., Rosa, M.L., Polyvyanyy, A.: Split miner: automated discovery of accurate and simple business process models from event logs. Knowl. Inf. Syst.  \textbf{59}(2),  251--284 (2019)

\bibitem{DBLP:journals/jal/BenderFPSS05}
Bender, M.A., Farach{-}Colton, M., Pemmasani, G., Skiena, S., Sumazin, P.: Lowest common ancestors in trees and directed acyclic graphs. J. Algorithms  \textbf{57}(2),  75--94 (2005)

\bibitem{pm4py}
Berti, A., van Zelst, S., Schuster, D.: Pm4py: A process mining library for python. Software Impacts  \textbf{17},  100556 (2023)

\bibitem{DBLP:conf/otm/BuijsDA12}
Buijs, J.C.A.M., van Dongen, B.F., van~der Aalst, W.M.P.: On the role of fitness, precision, generalization and simplicity in process discovery. In: On the Move to Meaningful Internet Systems: {OTM} 2012, Confederated International Conferences: CoopIS, DOA-SVI, and {ODBASE} 2012, Rome, Italy, September 10-14, 2012. Proceedings, Part {I}. Lecture Notes in Computer Science, vol.~7565, pp. 305--322. Springer (2012)

\bibitem{BPMN}
Chinosi, M., Trombetta, A.: {BPMN:} an introduction to the standard. Comput. Stand. Interfaces  \textbf{34}(1),  124--134 (2012)

\bibitem{DBLP:conf/bpmn/DumasGP10}
Dumas, M., Garc{\'{\i}}a{-}Ba{\~{n}}uelos, L., Polyvyanyy, A.: Unraveling unstructured process models. In: Business Process Modeling Notation - Second International Workshop, {BPMN} 2010, Potsdam, Germany, October 13-14, 2010. Proceedings. Lecture Notes in Business Information Processing, vol.~67, pp.~1--7. Springer (2010)

\bibitem{DBLP:conf/caise/DumasRMMRS12}
Dumas, M., Rosa, M.L., Mendling, J., M{\"{a}}esalu, R., Reijers, H.A., Semenenko, N.: Understanding business process models: The costs and benefits of structuredness. In: Advanced Information Systems Engineering - 24th International Conference, CAiSE 2012, Gdansk, Poland, June 25-29, 2012. Proceedings. Lecture Notes in Computer Science, vol.~7328, pp. 31--46. Springer (2012)

\bibitem{IM}
Leemans, S.J.J., Fahland, D., van~der Aalst, W.M.P.: Discovering block-structured process models from event logs - {A} constructive approach. In: Application and Theory of Petri Nets and Concurrency - 34th International Conference, {PETRI} {NETS} 2013, Milan, Italy, June 24-28, 2013. Proceedings. Lecture Notes in Computer Science, vol.~7927, pp. 311--329. Springer (2013)

\bibitem{DBLP:conf/bpm/LeemansFA13}
Leemans, S.J.J., Fahland, D., van~der Aalst, W.M.P.: Discovering block-structured process models from event logs containing infrequent behaviour. In: Business Process Management Workshops - {BPM} 2013 International Workshops, Beijing, China, August 26, 2013, Revised Papers. Lecture Notes in Business Information Processing, vol.~171, pp. 66--78. Springer (2013)

\bibitem{7PMG}
Mendling, J., Reijers, H.A., van~der Aalst, W.M.P.: Seven process modeling guidelines {(7PMG)}. Inf. Softw. Technol.  \textbf{52}(2),  127--136 (2010)

\bibitem{petri_net}
Weber, M., Kindler, E.: The petri net markup language. In: Petri Net Technology for Communication-Based Systems - Advances in Petri Nets. vol.~2472, pp. 124--144. Springer (2003)

\bibitem{DBLP:journals/is/WeerdtBVB12}
Weerdt, J.D., Backer, M.D., Vanthienen, J., Baesens, B.: A multi-dimensional quality assessment of state-of-the-art process discovery algorithms using real-life event logs. Inf. Syst.  \textbf{37}(7),  654--676 (2012)

\bibitem{FHM}
Weijters, A.J.M.M., Ribeiro, J.T.S.: Flexible heuristics miner {(FHM)}. In: Proceedings of the {IEEE} Symposium on Computational Intelligence and Data Mining, {CIDM} 2011, part of the {IEEE} Symposium Series on Computational Intelligence 2011, April 11-15, 2011, Paris, France. pp. 310--317. {IEEE} (2011)

\bibitem{DBLP:conf/cidm/WeijtersR11}
Weijters, A.J.M.M., Ribeiro, J.T.S.: Flexible heuristics miner {(FHM)}. In: Proceedings of the {IEEE} Symposium on Computational Intelligence and Data Mining, {CIDM} 2011, part of the {IEEE} Symposium Series on Computational Intelligence 2011, April 11-15, 2011, Paris, France. pp. 310--317. {IEEE} (2011)

\end{thebibliography}

\end{document}